\documentclass[peerreview,12pt,twocolumn,double,draftclsnofoot,romanappendices]{IEEEtran}
\usepackage[cmex10]{amsmath}
\usepackage{graphics}
\usepackage{mdwmath,mdwtab,url}
\usepackage[caption=false,font=footnotesize]{subfig}
\usepackage{stfloats}
\usepackage{nomencl}
\usepackage{amssymb}
\usepackage{rotating}
\usepackage{psfrag}
\usepackage{color}
\usepackage{hyperref}
\usepackage{cite}
\usepackage{epstopdf}

\newtheorem{theorem}{ {Theorem}}
\newtheorem{corollary}{ {Corollary}}

\newtheorem{definition}{{Definition}}

\newtheorem{remark}{ {Remark}}
\newcommand{\mc}{\mathcal} 

\newcommand{\tc}{\textcolor}
\definecolor{orange}{rgb}{1,0.5,0.0}
\begin{document}
%
 
\title{Multiaccess Channel with Partially Cooperating Encoders and Security Constraints}

 \author{Zohaib~Hassan~Awan, Abdellatif~Zaidi, and Luc~Vandendorpe,~\IEEEmembership{Fellow,~IEEE} 
				
\thanks{Copyright (c) 2013 IEEE. Personal use of this material is permitted. However, permission to use this material for any other purposes must be obtained from the IEEE by sending a request to pubs-permissions@ieee.org.}

\thanks{Zohaib Hassan Awan and Luc Vandendorpe are with the ICTEAM institute (\'{E}cole Polytechnique de Louvain), Universit\'e catholique de Louvain, Louvain-la-Neuve 1348, Belgium. Email: \{zohaib.awan,luc.vandendorpe\}@uclouvain.be}
 
\thanks{Abdellatif Zaidi is with the Universit\'e Paris-Est Marne-la-Vall\'ee, 77454 Marne-la-Vall\'ee Cedex 2, France. Email: abdellatif.zaidi@univ-mlv.fr}

\thanks{This work has been supported in part by the European Commission in the framework of the Network of Excellence in Wireless Communications (NEWCOM\#), and the Concerted Research Action, SCOOP. The authors would also like to thank BELSPO for the support of the IAP BESTCOM  network.}


\thanks{ The result in this work was presented in part at the 50th Annual Allerton Conference on Communication, Control, and Computing, Monticello, IL, USA, Oct. 2012.}
}

\markboth{To Appear in IEEE Transactions on Information Forensics and Security }{}

\maketitle
 
\begin{abstract}
We study a special case of Willems's two-user multi-access channel with partially cooperating encoders from a security perspective. This model differs from Willems's setup in that only one encoder, Encoder 1, is allowed to conference; Encoder 2 does not transmit any message, and there is an additional passive eavesdropper from whom the communication should be kept secret. For the discrete memoryless (DM) case, we establish inner and outer bounds on the capacity-equivocation region. The inner bound is based on a combination of Willems's coding scheme, noise injection and additional binning that provides randomization for security. For the memoryless Gaussian model, we establish lower and upper bounds on the secrecy capacity. We also show that, under certain conditions, these bounds agree in some extreme cases of cooperation between the encoders.  We illustrate our results through some numerical examples.
\end{abstract}


\begin{IEEEkeywords}
Multi-access channel, wire-tap channel, conferencing, eavesdropping, security.
\end{IEEEkeywords}


 %
\section{Introduction}
 
\IEEEPARstart{T}{raditionally} security in communication networks is achieved through encryption algorithms, implemented in the upper layers of the protocol stack. Wyner introduced a basic information-theoretic model to study security by exploiting the physical layer attributes of the channel \cite{wyner}. The wiretap channel studied by Wyner consists of a source, a destination (legitimate receiver) and an eavesdropper. The source communicates with the destination, and wishes to conceal the messages that it sends from the eavesdropper. Wyner establishes the secrecy capacity of this model, i.e., the maximum amount of information that can be sent from the source to the destination while leaking absolutely no information to the eavesdropper, in the discrete memoryless (DM) case when the source-to-eavesdropper channel is a degraded version of the source-to-destination channel. The secrecy capacity of the memoryless Gaussian version of Wyner's wiretap model is obtained in \cite{leung}. In \cite{csiszar}, Csisz\'{a}r and K\"orner generalize Wyner's wiretap model to a broadcast model with confidential messages (BCC). In this model, the source communicates with two destinations; it sends two messages, a common message that is intended to be decoded by both destinations as well as an individual message that is intended to be decoded by only one destination and be kept secret from the other destination. For the transmission of the individual message, the destination that recovers only the common message then plays the role of an eavesdropper. Csisz\'{a}r and K\"orner characterize the capacity-equivocation region and the secrecy capacity region of the studied broadcast model with confidential messages.

The seminal work by Wyner has been extended to a variety of models, including the parallel broadcast channel with confidential messages \cite{liangshamai}, the multi-antenna wiretap channel \cite{khisti,oggier,Tie}, the multi-access wiretap channel \cite{tekin,liangpoor,tekin2}, the relay-eavesdropper channel   \cite{yuksel1,yuksel2,xianghe,lai,vaneet}, the parallel relay-eavesdropper channel \cite{zohaibj}, the interference channel with confidential messages \cite{tang,onur,ruoheng,vilela}  and the fading wiretap channel \cite{barros,gopala}. The reader may refer to \cite{liangbook} for recent advances on aspects related to information-theoretic security.
 
In this work, we investigate the problem of secure communication over a multi-access channel (MAC) with partially cooperating encoders. The MAC with partially cooperating encoders and no security constraints was studied by Willems in \cite{willems}. In this model, prior to transmitting their respective messages, the two encoders are allowed to communicate with each other over noiseless bit-pipes of finite-capacities. Willems characterizes the complete capacity region of this model for the DM case. In \cite{wigger}, Bross \textit{et al.} establish the capacity region of the memoryless Gaussian version of Willems's model. In both \cite{willems} and \cite{wigger}, among other observations, it is shown in particular that holding a conference prior to the transmission, enlarges the capacity region relative to the standard MAC with independent inputs.
 
\begin{figure}
\psfragscanon
\begin{center}
\psfrag{W}[c][c]{$W$}
\psfrag{C}[][l]{$C_{12}$}
\psfrag{Y}[c][c]{$Y^n$}
\psfrag{Z}[c][c]{\tc{red}{$Z^n$}}
\psfrag{E}[c][c][0.9]{Encoder 2}
\psfrag{F}[c][c][0.9]{Encoder 1}
\psfrag{I}[c][c]{$X_2^n$}
\psfrag{J}[c][c]{$X_1^n$}
\psfrag{U}[c][c]{\:\:$\hat{W}$}
\psfrag{G}[c][c][0.9]{$p(y,z|x_1,x_2)$}
\psfrag{M}[c][c][0.8]{\bf{Helping Encoder}}
\psfrag{K}[c][c]{$\bf{MAC}$}
\psfrag{Q}[c][c][0.8]{\bf{Source}}
\psfrag{D}[c][c][0.9]{Decoder}
\psfrag{X}[c][c][0.88]{Eavesdropper}
\psfrag{O}[c][c][0.8]{\bf{Legitimate Receiver}}
\psfrag{V}[c][c]{\tc{red}{$W$}}
\includegraphics[width=\linewidth]{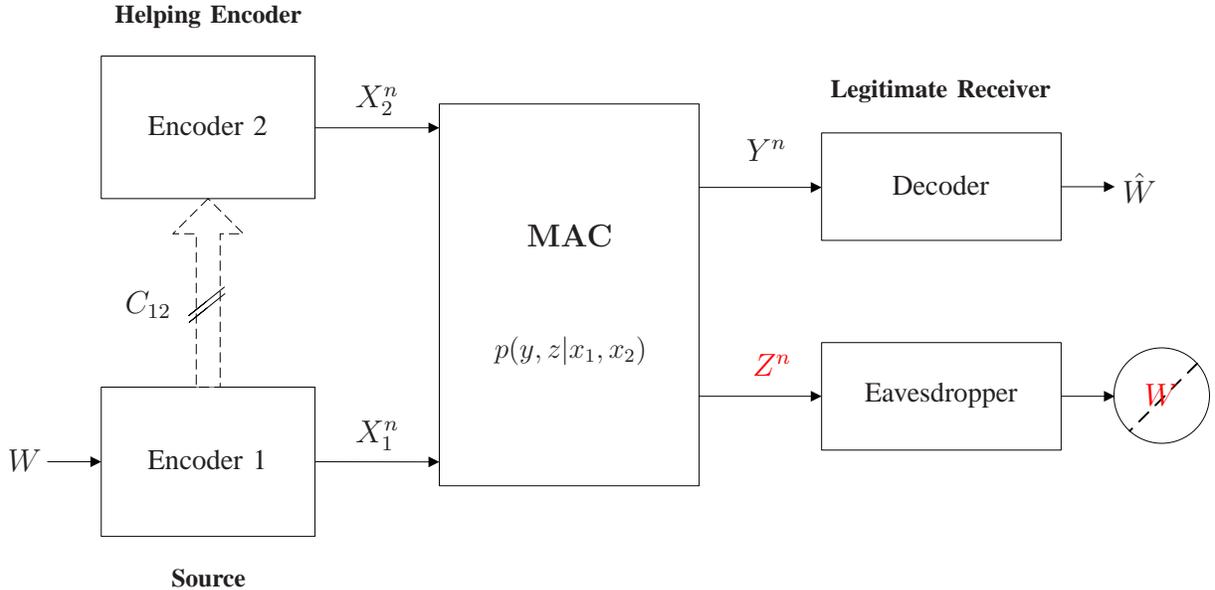}
\end{center}
\caption{Multi-access channel with partially cooperating encoders and security constraints.}
\psfragscanoff
 \label{DMC}
\end{figure}

We study a special case of Willems's setup with an additional security constraint on the communication. More specifically, as shown in Figure~\ref{DMC}, we consider a two-user multi-access channel in which the two users can   cooperate partially through a unidirectional noiseless bit-pipe of finite capacity $C_{12}$. Also, we restrict the role of Encoder 2 to only helping Encoder 1, i.e., Encoder 2 has no message of its own to transmit. Furthermore, we assume that there is a passive eavesdropper who overhears the transmission and from whom the communication should be kept secret. The eavesdropper is passive in the sense that it only listens to the transmitted information without modifying it. The role of Encoder 2 is then to only help Encoder 1 communicate with the legitimate receiver while keeping the transmitted information \textit{secret} from the eavesdropper. From a practical viewpoint, this model may be appropriate for example for the study of the role of backbone connections among base stations for securing transmission in cellular environments. In this paper, we study the capacity-equivocation region of this model.

The MAC model that we study in this paper has some connections with a number of related works studied previously. Compared with the orthogonal relay-eavesdropper channel studied in \cite{vaneet}, the orthogonal link between the source and the relay is replaced here by a noiseless bit-pipe of finite capacity $C_{12}$. Compared with the wiretap channel with a helper interferer (WT-HI) studied in \cite{tang}, our model permits cooperation among the encoders. Finally, compared with the primitive relay channel of \cite{kim}, our model imposes security constraints on the transmitted message.

\vspace{.5em}
\textit{\textbf{Contributions.}} Our main contributions in this paper can be summarized as follows. For the DM case, we establish outer and inner bounds on the capacity-equivocation region. The coding scheme that we use for the inner bound is based on an appropriate combination of Willems coding scheme \cite{willems}, noise injection \cite[Theorem 3]{lai} and binning  for randomization.  We obtain our converse proof by extending the converse proof of \cite{willems} to account for the security constraint and that of \cite{csiszar} to account for the unidirectional noiseless bit-pipe cooperation among the encoders. In doing so, we show that one needs to redefine the involved auxiliary random variables appropriately. We note that characterizing the capacity-equivocation region of our model in the general setting is not easy; and, in fact, the capacity-equivocation region or secrecy-capacity of closely related models that are reported in the literature, such as \cite{tang,xainghe,LYT08}, are still to be found -- the model of \cite{tang} can be seen as a special case of our model obtained by taking a noiseless bit-pipe of zero capacity. From this perspective, the inner and outer bounds that we develop here can be seen as one step ahead towards a better understanding of the full capacity-equivocation region of the model that we study in this paper.

Next, we study the Gaussian memoryless model. In this case, we focus only on perfect secure transmission. For this model, we establish lower and upper bounds on the secrecy capacity. The coding scheme that we use to establish the lower bound uses ideas that are essentially similar to those for the DM case. The upper bound on the secrecy capacity does not involve auxiliary random variables and, so, is computable. Furthermore, it has the same expression as the secrecy capacity of the Gaussian wiretap channel with a two-antenna transmitter, single-antenna legitimate receiver and single-antenna eavesdropper \cite{khisti,oggier,Tie}.

We show that our lower bound performs well in general and is optimal in some extreme cases of cooperation among the encoders, including when the two encoders  fully cooperate, i.e., $C_{12} := \infty$.  For the case in which the two encoders do not conference, i.e., $C_{12}:=0$, the model that we study reduces to a wiretap channel with a helper interferer \cite{tang,xainghe}. In this case, our coding scheme reduces to merely injecting statistically independent noise \cite[Theorem 3]{lai}; and, by comparing it to the upper bound that we develop, we show that it is optimal under certain conditions. For the case of full cooperation among the encoders, i.e., $C_{12}:=\infty$, our coding scheme reduces to full two-antenna cooperation for providing secrecy in the context of multi-antenna wiretap channels \cite{khisti,oggier,Tie}.
 
The rest of the paper is organized as follows. Section II provides a formal description of the channel model. In section III, we study the DM setting, and establish inner and outer bounds on the capacity-equivocation region. In section IV, we establish lower and upper bounds on the secrecy-capacity for the memoryless Gaussian model, and study some extreme cases of cooperation among the encoders. We illustrate these results through some numerical examples in section V. Section VI concludes the paper.

\vspace{.5em}
\textit{\textbf{Notations.}} In this paper, the notation $X^n$ is used as a shorthand for $(X_1,X_2,\hdots,X_n)$, the notation $X_i^n$ is used as a shorthand for  $(X_i,X_{i+1},\hdots,X_n)$, the notation $ \mc X^n$ is used as a shorthand for $(\mc X_1 \times \mc X_2 \hdots \times \mc X_n)$, the notation $|\mc X|$ denotes the cardinality of set $\mc X$, $\mathbb{E}\{.\}$ denotes the expectation operator, $\mc N(\mu,\sigma^2)$ denotes the Gaussian distribution with $\mu$-mean and $\sigma^2$-variance, the boldface letter $\bf{X}$ denotes the covariance matrix; $H(\cdot)$, $h(\cdot)$ denote the entropy of the discrete and continuous random variables respectively and $I(X;Y)$ defines the mutual information between random variable $X$ and $Y$. We define the functions $\mc{C}(x)=\log_2(1+x)$ and $[x]^+ = \max\{0,x\}$. Throughout the paper the logarithm function is taken to the base 2.
 
\section{Channel Model and Definitions}
 
Consider the model depicted in Figure~\ref{DMC}. Encoder 1 wishes to send a confidential message to the legitimate receiver, in the presence of a passive eavesdropper that overhears the transmitted information and cannot modify it. In doing so, Encoder 1 can get help from a second encoder, Encoder 2, to whom it is connected through a noiseless bit pipe of finite capacity $C_{12}$. Encoder 2 has no message of its own to transmit, and is dedicated entirely to help Encoder 1 conceal its message from the eavesdropper. The eavesdropper is assumed to be of unlimited computational complexity and is fully informed about the codebooks used at the encoders.

More formally, let $W$ denote the message to be transmitted, taken uniformly from the set $\mc W = \{1,\hdots,2^{nR}\}$. Encoder 1 is allowed to conference the message $W$ to Encoder 2 using $K$ communicating functions $\{ \phi_{11}, \phi_{12},\hdots,\phi_{1K}\}$, over the noiseless bit-pipe. Let $G_{1k}:=  \phi_{1k}(W)$, defined as the output of the communication process for the k-$th$ communication, where $G_{1k}$ ranges over the finite alphabet $\mc G_{1k}$,   $k=1,\hdots,K$. The information conferenced is bounded due to the finiteness of noiseless bit-pipe capacity between the two encoders. A conference is permissible if communication functions are such that
\begin{equation}
\label{defineC12}
\sum_{k=1}^K \log{|\mc {G}_{1k}|} \le nC_{12}.
\end{equation}
To transmit the message $W$, Encoder 1 sends a codeword $X_1^n \in \mc X^n_1$,  where $\mc X_1$ designates the input alphabet at Encoder 1. Encoder 2 transmits a codeword $X_2^n \in \mc X^n_2$ where $\mc X_2$ designates the input alphabet at Encoder 2. Let $\mc Y$ and $\mc Z$ designate the output alphabets at the legitimate receiver and eavesdropper, respectively. The legitimate receiver gets the channel output $Y^n \in \mc Y^n$, and tries to estimate the transmitted message from it. The eavesdropper overhears the channel output $Z^n \in \mc Z^n$. The transmission over the channel is characterized by the memoryless conditional probability $p(y,z|x_1,x_2)$. The channel is memoryless in the sense that
\begin{eqnarray}
\label{channel_dmc}
p(y^n,z^n|x^n_1,x^n_2) = \prod_{i=1}^{n} p(y_i,z_{i} | x_{1,i},x_{2,i}).
\end{eqnarray}

\begin{definition}
A $(2^{nR},n)$ code for the multi-access model with partially cooperating encoders shown in Figure~\ref{DMC} consists of encoding functions\footnote{The source encoder, $\phi_1$, and helper encoder, $\phi_2$,  are stochastic encoders that introduce additional randomization to increase secrecy.}
\begin{align}
\phi_1 \:\: &: \:\: \mc W \longrightarrow \mc X^n_1 ,\nonumber\\
\phi_{1k} \:\: &: \:\: \mc W \longrightarrow \mc{G}_{1k}, \:\:\:\:    k=1,...,K,\nonumber\\
\phi_2 \:\: &: \:\: \{1,\hdots,2^{nC_{12}}\} \longrightarrow \mc X^n_2,
\end{align}
and a decoding function $\psi(\cdot)$ at the legitimate receiver
\begin{align}
\psi \:\: &: \:\: \mc Y^n  \longrightarrow \mc W.
\end{align}
\end{definition}
\vspace{.5em}

\begin{definition}
 The average error probability for the $(2^{nR},n)$ code is defined as
\begin{eqnarray}
P_{e}^n = \frac{1}{2^{n{R}}} \sum_{W \in \mathcal{W}}\text{Pr}\{ \hat{W} \ne W | W  \}.
\end{eqnarray}
\end{definition}
\vspace{.5em}

\noindent The eavesdropper overhears to what the encoders transmit and tries to guess the information from it. The equivocation rate per channel use is defined as $R_e = H(W| Z^n)/n$.

\vspace{.5em}
\begin{definition}
A rate-equivocation pair $({{R},{R}_{e}})$ is said to be achievable if for any $\epsilon > 0$ there exists a sequence of codes $(2^{nR},n)$ such that for any $n \ge n(\epsilon)$
\setlength{\arraycolsep}{0.2em}
\begin{eqnarray}
\label{def4}
\frac{H(W)}{n} &\ge& R - \epsilon,\notag \\
\frac{H(W| Z^n) }{n} &\ge& R_e - \epsilon , \notag \\
P_{e}^n &\le& \epsilon.
\end{eqnarray}
\end{definition}

\vspace{.5em}
\begin{definition}
The secrecy capacity is defined as the maximum achievable rate at which the communication rate is equal to the equivocation rate, i.e., $(R,R_e)= (R,R)$.
\end{definition}
\setlength{\arraycolsep}{5pt}

\section{Discrete Memoryless Case}

In this section we establish outer and inner bounds on the capacity-equivocation region for the MAC with partially cooperating encoders shown in Figure~\ref{DMC}.

\subsection{Outer Bound}
The following theorem provides an outer bound on the capacity-equivocation region of the MAC with partially cooperating encoders and security constraints shown in Figure~\ref{DMC}.
\begin{theorem}
\label{outer}
For the MAC with partially cooperating encoders and security constraints shown in Figure~\ref{DMC}, and for any achievable rate-equivocation pair $(R,R_e)$, there exist some random variables $U\leftrightarrow(V_{1},V_{2})\leftrightarrow(X_{1},X_{2})\leftrightarrow(Y,Z)$, such that $(R,R_e)$ satisfies


 \setlength{\arraycolsep}{0.2em}
\begin{align}
\label{Upeq}
R  &\leq \min  \{ I(V_1,V_2; Y),\: I(V_1 ; Y | V_2 )+ C_{12} \} \nonumber\\
R_e &\leq R \nonumber\\
R_e &\le \min \{ I(V_1,V_2;Y|U) -  I(V_1,V_2; Z | U), I(V_1; Y | V_2, U)+C_{12}  -  I(V_1,V_2; Z | U)\}.
 \end{align}
\end{theorem}

 
\vspace{.5em}
\begin{IEEEproof}
 The proof of Theorem~\ref{outer} is given in Appendix~\ref{appendix_outer}.
\end{IEEEproof}
\vspace{.5em}
\begin{remark}
The proof of Theorem~\ref{outer} extends the converse proof of \cite{csiszar} to the case of two encoders, and extends the converse proof of \cite{willems} so as to account for the imposed security constraint. Furthermore, the outer bound of Theorem \ref{outer} reduces to the secrecy capacity of Wyner's wiretap channel \cite{wyner} by removing the helping encoder, Encoder 2.
\end{remark}
 
\vspace{0.5em}
\begin{remark}
In the special case in which $C_{12}:=0$,  the model in Figure~\ref{DMC} reduces to a transmitter (Encoder 1) sending a confidential message to its intended receiver in the presence of a passive eavesdropper and with the help of an external independent interferer (Encoder 2). This model is referred to as being a wiretap channel with a helping interferer (WT-HI), and is studied in \cite{tang,xainghe}. The capacity-equivocation region of the WT-HI is still unknown. In \cite{tang}, and also \cite{xainghe}, the authors derive achievable secrecy rates as well as computable upper bounds on the secrecy capacity of the WT-HI. The outer bounds of \cite{tang,xainghe} are of Sato-type. By specializing the outer bound of Theorem~\ref{outer} to the case $C_{12}:=0$, one readily obtains an alternative outer bound on the capacity equivocation region of the WT-HI. It is not easy to compare the obtained outer bound with the outer bounds of \cite{tang,xainghe}, since the former involves auxiliary random variables. 
\end{remark}

\subsection{Inner Bound}
We now turn to establish an inner bound on the capacity-equivocation region of the MAC with partially cooperating encoders and security constraints shown in Figure~\ref{DMC}.  The following theorem states the result.

\vspace{0.5em}
\begin{theorem}
\label{Th_inner}
For the MAC with partially cooperating encoders and security constraints shown in Figure \ref{DMC}, the rate pairs in the closure of the convex hull of all $(R,R_e)$ satisfying
 
\setlength{\arraycolsep}{0.2em}
\begin{align}
\label{inner}
R   &\le \min  \{ I(V_1, V_2 ; Y|U),\: I(V_1 ; Y | V_2, V, U) + C_{12} \} \nonumber\\
R_e &\leq {R} \nonumber\\
R_e &\leq [\min\{ I(V_2 ; Y|V,U), I(V_2; Z|V_1,V,U)\} +\min\{I(V_1, V_2 ; Y|U),I(V_1 ; Y | V_2, V, U) + C_{12}\}\notag\\& \hspace{1.3em}-  I(V_1, V_2; Z | U)]^+
\end{align}


\noindent for some measure $p(u,v,v_{1},v_{2},x_{1},x_{2},y,z)=p(u)p(v|u)p(v_1|v,u)p(v_2|v,u)p(x_{1}|v_{1})p(x_{2}| v_{2})$\\$.p(y,z|x_{1},x_{2})$, are achievable.
 \setlength{\arraycolsep}{5pt}
\end{theorem}
\vspace{0.5em}
\begin{IEEEproof}[Outline of Proof] \\
We briefly outline the coding scheme that we use to prove the achievability of the inner bound of Theorem \ref{Th_inner}, and relegate  the details of the proof to Appendix \ref{appendix_inner}. The inner bound of Theorem \ref{Th_inner} is based on a coding scheme that consists in appropriate careful combination of Willems's coding scheme \cite{willems}, noise injection \cite[Theorem 3]{lai} and binning for randomization to provide security. Let $W$ denote the message to be transmitted. Using the noiseless bit-pipe of finite capacity, Encoder 1 conferences a part of the information message $W$ to Encoder 2. After completion of the conferencing process, this part can be regarded as a common information to be transmitted by both encoders. The random variable $V$ in Theorem \ref{Th_inner} represents this common information. The part of the information message that is sent only by Encoder 1 can be regarded as an individual message. The random variable $V_1$ in Theorem \ref{Th_inner} represents this individual information. The input of Encoder 2 is composed of the common information, which it has received through noiseless finite capacity link from Encoder 1, and a statistically independent artificial noise component. The random variable $V_2$ in Theorem \ref{Th_inner} represents the input from Encoder 2. The transmission of both common information and artificial noise components at Encoder 2  in Theorem~\ref{Th_inner} is adjusted by appropriate selection of random variable $V$. Additional random binning is employed to secure both individual and common information from the passive eavesdropper \cite{wyner}. Finally, the random variable $U$ in Theorem \ref{Th_inner} stands for a channel prefix.
\end{IEEEproof}
\vspace{0.5em}
\begin{remark}
The region established in Theorem~\ref{Th_inner} reduces to the special case $R_2=0$ of the capacity region of the MAC with cooperating encoders and no security constraints in \cite[Theorem 1]{willems} by setting $R_e:=0$, $U:=$constant, $V_1:=X_1$ and $V=V_2=X_2$ in \eqref{inner}.
\end{remark}
\vspace{0.5em}
\begin{remark}
As we indicated previously, in the special case in which $C_{12}:=0$,  the model of Figure~\ref{DMC} reduces to a wiretap channel with a helping interferer (WT-HI). By setting $R_e:=R$ (i.e., restricting to the case of perfect secrecy) and $U = V =\phi$ in \eqref{inner}, we obtain the following lower bound on the secrecy capacity of the WT-HI,
\begin{eqnarray}
\label{c12}
R_e \leq \max\: [\min\{I(V_1, V_2 ; Y)-I(V_1, V_2; Z ), I(V_1 ; Y | V_2) -  I(V_1; Z )\}]^+
\label{lower-bound-secrecy-capacity-WTHI}
\end{eqnarray}
where the maximization is over joint measures of the form $p(v_1)p(v_2)p(x_{1}|v_{1})p(x_{2}| v_{2})$. In \cite{tang}, the authors establish several achievable secrecy rates for the WT-HI for different regimes of the relative strength of the interference. The lower bound \ref{lower-bound-secrecy-capacity-WTHI} has an expression that is essentially similar to one that is developed for the case of a strong interference regime in \cite[Section III-C]{tang}; but is potentially larger since it involves auxiliary random variables $V_1$ and $V_2$ in place of the inputs $X_1$ and $X_2$ in \cite[Section III-C]{tang}. The specific choice $V_1:=X_1$ and $V_2:=X_2$ gives the lower bound of \cite[Section III-C]{tang} in the case of strong interference.
\end{remark}

\section{Memoryless Gaussian Model}

In this section, we study the Gaussian version of the MAC with partially cooperating encoders and security constraints shown in Figure \ref{DMC}. We only focus on the case of perfect secrecy.
\subsection{Channel Model}
For the Gaussian model, the outputs of the MAC at the legitimate receiver and eavesdropper for each symbol time are given by
\begin{align}
\label{gchan}
Y  &= h_{1d}{X_{1 }}+ h_{2d} X_{2 }+N_{1 }\notag\\
Z  &= h_{1e}{X_{1 }}+h_{2e} X_{2 }+N_{2 }
\end{align}
where $ h_{1d}$, $h_{2d}$, $h_{1e}$, and $h_{2e}$ are the channel gain coefficients associated with Encoder 1-to-destination (1-D),  Encoder 2-to-destination (2-D), Encoder 1-to-eavesdropper (1-E), and Encoder 2-to-eavesdropper (2-E) links respectively. The noise processes $\{N_{1,i}\}$ and $\{N_{2,i}\}$ are independent and identically distributed (i.i.d) with the components being zero mean Gaussian random variables with variances $\sigma_1^2$ and $\sigma_2^2$, respectively; and $X_{1,i}$ and $X_{2,i}$ are the channel inputs from Encoder 1 and Encoder 2 respectively. The channel inputs are bounded by average block power constraints
\begin{align}
\label{p_con1}
&\sum_{i=1}^n \mathbb{E}[X_{1,i}^2] \le  nP_1, \qquad \sum_{i=1}^n \mathbb{E}[X_{2,i}^2]\le nP_2.
\end{align}

\subsection{Upper Bound on the Secrecy Capacity}
In this section, we establish an upper bound on the secrecy capacity on Gaussian MAC \eqref{gchan}.  We establish a computable upper bound using the techniques developed earlier to establish the secrecy capacity of a multiple-input multiple-output (MIMO) wiretap channel \cite{khisti,oggier,Tie} --- taking a setup with two antennas at the transmitter, one antenna at the legitimate receiver and one antenna at the eavesdropper in our case.
\vspace{.5em}
\begin{corollary}
\label{gaussian_upper_bound}
For the Gaussian MAC with partially cooperating encoders and security constraints \eqref{gchan}, an upper bound on the secrecy capacity is given by
\begin{align}
\label{guu}
R_e^{\text{up}} = \max_{\psi} [I(X_{1},X_{2};Y)-I(X_{1},X_{2};Z)]
\end{align}
where  $[X_1,X_2] \sim \mc{N}(\bf{0},\bf{K}_{P})$ with ${{\mc{K}}_{P}}  = \Big \{ {\bf{K}_{P}} : {\bf{K}_{P}}$=  $\left [
 \begin{smallmatrix}
  P_{1} & \psi\sqrt{P_{1} P_{2}}\\
\psi\sqrt{P_{1} P_{2}} & P_{2}
 \end{smallmatrix} \right ]$, $-1\le \psi \le {1} \Big\}$, with $\mathbb{E}[X^2_{1}]$,  $\mathbb{E}[X^2_{2}]$ satisfying \eqref{p_con1}.
\end{corollary}
 \vspace{.5em}
Alternatively, we can also establish the upper bound \eqref{guu} from the rate-equivocation region established for the DM case in Theorem \ref{outer}, as follows. Taking the first term of minimization in the bound on the equivocation rate in Theorem \ref{outer}, we get
\begin{eqnarray}
\label{nupper}
R_e  \le  \:\: \max \:\: [I(V_1,V_2;Y| U)-I(V_1,V_2 ;Z| U)]
\end{eqnarray}
where ${U}\leftrightarrow({V_1,V_2})\leftrightarrow({X_1,X_2})\leftrightarrow(Y,Z)$. The rest of the proof closely follows the bounding technique established in \cite{zohaibj}, in the context of a parallel relay-eavesdropper channel. More specifically, continuing from \eqref{nupper} we get

 
\setlength{\arraycolsep}{2pt}
\begin{eqnarray}
R_e &\leq&  I(V_{1},V_{2} ;Y| U)-I(V_{1},V_{2} ;Z| U)\notag\\
  &\overset{(a)}{\le}&  I(V_{1},V_{2} ;Y)- I(V_{1},V_{2} ;Z)\notag \\
    &{\le}&  I(V_1,V_2;Y,Z )-I(V_{1},V_{2} ;Z)\notag \\
   &\overset{(b)}{=}& [I( X_1,X_2;Y, Z )-I( X_1, X_2;Y, Z| V_1, V_2 )]- [I(X_1, X_2 ;Z)-I(X_1,X_2 ;Z| V_1, V_2 )]\notag\\
   &{=}& [I( X_1 ,X_2;Y,Z )-I(X_1, X_2 ;Z)] - [I( X_1, X_2;Y, Z| V_1 ,V_2 )-I(X_1, X_2 ;Z| V_1,V_2 )]\notag \\
     &{\le}& I( X_1, X_2;Y,Z )-I(X_1, X_2 ;Z) \notag\\
     &=&  I( X_1 ,X_2;Y | Z )
     \label{ubound}
     \end{eqnarray}

\noindent where $(a)$ follows  from the fact that the difference of conditional mutual information  $I(V_{1},V_{2} ;Y|U)-I(V_{1},V_{2} ;Z|U)$ is maximized by $U:=\text{constant}$ and $(b)$ holds since $({V_1,V_2})\leftrightarrow({X_1,X_2})\leftrightarrow(Y,Z)$ is a Markov chain.

\noindent Now, the upper bound in \eqref{ubound} can be tightened by using an argument previously used in \cite{khisti,oggier} in the context of multi-antenna wiretap channel. Noticing that the upper bound \eqref{nupper} depends on $p(y,z|x_1,x_2)$ only through its marginals $p(y|x_1,x_2)$ and $p(z|x_1,x_2)$, the upper bound \eqref{ubound} can be further tightened as
  \setlength{\arraycolsep}{5pt}
\begin{eqnarray}
\label{nuppern}
{R}_e \le    \min_{p(y',z'|x_1,x_2)} \max_{p(x_1,x_2)} I( X_1, X_2;Y' | Z' )
\end{eqnarray}
where the joint conditional  $p(y',z'|x_1,x_2)$ has the same marginals  as  $p(y,z|x_1,x_2)$, i.e.,  $p(y'|x_1,x_2)=p(y|x_1,x_2)$ and $p(z'|x_1,x_2)=p(z|x_1,x_2)$.

Following \cite{khisti,oggier}, it can be shown that the bound in \eqref{nuppern} is maximized with the jointly Gaussian inputs $[X_1,X_2] \sim \mc{N}(\bf{0},\bf{K}_{P})$, with $\mathbb{E}[X^2_{1}]$ and $\mathbb{E}[X^2_{2}]$ satisfying \eqref{p_con1}.
 
\noindent Finally, evaluation of the upper bound \eqref{nuppern} with these jointly Gaussian inputs and then the minimization over all possible correlations between $Y'$ and $Z'$ yield the desired result.
\vspace{.5em}
\subsection{Lower Bound on the Secrecy Capacity}

For the Gaussian MAC with partially cooperating encoders and security constraints \eqref{gchan}, we obtain a lower bound on the secrecy capacity by using our result for the DM model in Theorem \ref{Th_inner}. The results established for the DM case can be readily extended to memoryless channels with discrete time and continuous alphabets using standard techniques \cite[Chapter 7]{gallager}.
\vspace{.5em}
\begin{corollary}
\label{cor1}
For the Gaussian MAC with partially cooperating encoders and security constraints \eqref{gchan}, a lower bound on the secrecy capacity is given by

 
{\small{
\begin{align}
\label{lower_gaussian}
R_{e}^{\textrm{low}} =&
{\max_{
\substack{
0 \le \alpha \le 1,\\
0 \le \beta \le 1
}
}}
\bigg[\min \bigg \{\mathcal{C}\bigg(\frac{\beta|h_{2d}|^2 P_2}{\sigma_1^2+\alpha|h_{1d}|^2 P_1}\bigg),
\mathcal{C}\bigg(\frac{\beta|h_{2e}|^2 P_2}{\sigma_2^2}\bigg) \bigg\} \notag\\&\quad\quad+
\min \bigg \{\mathcal{C}\bigg(\frac{\alpha|h_{1d}|^2 P_1}{\sigma_1^2}\bigg)+C_{12},
\mathcal{C}\bigg(\frac{|h_{1d}|^2P_1+ |h_{2d}|^2P_2+2\sqrt{\bar{\alpha}\bar{\beta}|h_{1d}|^2P_1|h_{2d}|^2P_2}}{{\sigma_1}^2}\bigg)\bigg\} \notag\\&\quad\quad- \mathcal{C}\bigg(\frac{|h_{1e}|^2P_1+|h_{2e}|^2P_2+2\sqrt{\bar{\alpha}\bar{\beta}|h_{1e}|^2P_1|h_{2e}|^2P_2}}{{\sigma}_{2}^2}\bigg)\bigg]^+.
\end{align}
}}

\setlength{\arraycolsep}{5pt}
\normalsize
\end{corollary}
\vspace{.5em}
\begin{IEEEproof}
The achievability follows by computing the inner bound in Theorem \ref{Th_inner} with the choice $U:=\text{constant}$, $V_1:=X_1$ and $V_2:=X_2$, ${X_{1}}:=\sqrt{(\alpha P_1)}{\tilde{X}_{1}} + \sqrt{(\bar{\alpha} P_{1})}V$, ${X_{2}}:=\sqrt{(\beta P_2)}{\tilde{X}_{2}} + \sqrt{(\bar{\beta} P_{2})}V$, where $\tilde{X}_{1}$,  $\tilde{X}_{2}$ and $V$ be independent random variables with $\mc N(0,1)$, and $\alpha \in [0,1]$, $\bar{\alpha}:=1-\alpha$, $\beta \in [0,1]$, and $\bar{\beta}:=1-\beta$. Straightforward algebra that is omitted for brevity gives \eqref{lower_gaussian}.
\end{IEEEproof}
\vspace{.5em}

\normalsize
\subsection{Analysis of Some Extreme Cases}
In this section we study two special cases of the Gaussian MAC \eqref{gchan} with partially cooperating encoders shown in Figure \ref{DMC}, where the capacity of the bit-pipe is either,
\begin{enumerate}
\item $C_{12}$ = 0, or
\item $C_{12}$ = $\infty$.
\end{enumerate}
The first case corresponds to the wiretap channel with a helping interferer (WT-HI) studied in \cite{tang,xainghe} . The second case corresponds to a two-antenna transmitter wiretap channel \cite{km,oggier}.

\vspace{0.5em}
\subsubsection{Case $C_{12} := 0$}
\vspace{0.5em}

In this case the encoders do not cooperate. Since Encoder 2 does not know the common information to transmit, it only injects statistically independent artificial noise.
\vspace{0.5em}
\begin{corollary}
\label{corollary1}
For the Gaussian model \eqref{gchan} with $C_{12} := 0$:
 
\begin{enumerate}
\item An upper bound on the secrecy capacity is given by

 
\begin{align}
\label{equpbound}
R_e^{\textrm{up}}  =    \max_{\substack{
\mathbb{E}[X_1^2] \le P_1, \\
\mathbb{E}[X_2^2] \le P_2}}\bigg[ \mc{C} \Big(\frac{|h_{1d}|^2\mathbb{E}[X_1^2] }{\sigma_{1}^2}\Big) - \mc{C} \Big(\frac{|h_{1e}|^2\mathbb{E}[X_1^2] }{\sigma_{2}^2+|h_{2e}|^2\mathbb{E}[X_2^2] }\Big)\bigg]^+.
\end{align}

 

\item A lower bound on the secrecy capacity is given by
 
\begin{align}
\label{eqbound}
R_e^{\textrm{low}}  =    \max\: \bigg[ \mc{C} \Big(\frac{|h_{1d}|^2\mathbb{E}[X_1^2] }{\sigma_{1}^2}\Big) - \mc{C} \Big(\frac{|h_{1e}|^2\mathbb{E}[X_1^2] }{\sigma_{2}^2+|h_{2e}|^2\mathbb{E}[X_2^2] }\Big)\bigg]^+
\end{align}
 
 
 
 where the maximization is over $\mathbb{E}[X_1^2] \le P_1$ and $\mathbb{E}[X_2^2] \le P_2$ such that
\begin{eqnarray}
\label{condition}
 \mathcal{C}\Big(\frac{|h_{2d}|^2\mathbb{E}[X_2^2]}{|h_{1d}|^2\mathbb{E}[X_1^2]+{\sigma}_1^2}\Big) \ge   \mc{C} \Big(\frac{|h_{2e}|^2\mathbb{E}[X_2^2]}{\sigma_{2}^2}\Big).
\end{eqnarray}
\end{enumerate}
\end{corollary}
 
\vspace{0.5em}
\begin{IEEEproof}
\textit{\textbf{Upper Bound.}}
We bound the term in \eqref{equpbound} as follows. The proof follows by using elements from an upper bounding technique developed in \cite{vaneet}. We assume that there is a noiseless link between Encoder 2 and the legitimate receiver, and the eavesdropper is \textit{constrained} to treat Encoder 2's signal as unknown noise. The upper bound established for this model, with full cooperation between Encoder 2 and the legitimate receiver and a constrained eavesdropper, also applies to the model of Corollary~\ref{corollary1}.

With full cooperation between Encoder 2--legitimate receiver link, the legitimate receiver can remove the effect of Encoder 2 transmission from the output $Y$ of the MAC \eqref{gchan} (since the input from Encoder 2 is independent from Encoder 1 transmission because $C_{12}=0$). The equivalent channel model at the legitimate receiver is then given by
\begin{align}
Y'  &= h_{1,d}X_{1 }+N_{1 }.
\end{align}
The eavesdropper is constrained in the sense that it is \textit{restricted} not to decode Encoder 2 signals. For the constrained eavesdropper  Encoder 2's transmission acts as unknown noise, the worst case is obtained with the $X_2$  being Gaussian distributed \cite{vaneet}. The equivalent channel model at the eavesdropper is given by
\begin{align}
{Z'}&= h_{1,e}X_{1}+\underbrace{h_{2,e} X_{2} }_{\text{unknown noise}}+{N_{2}}.
\end{align}
The equivalent channel model, with full cooperation between Encoder 2-to-legitimate receiver link and worst case Encoder 2-to-constrained eavesdropper transmission, reduces to the Gaussian wiretap channel, the secrecy capacity of which is established in \cite{leung}, i.e.,
\begin{eqnarray}
\label{newup2}
C_s =   \max_{p(x_1)p(x_2)}I(X_{1};Y')-I(X_{1};Z')
\end{eqnarray}
where  $X_{1}\sim \mc{N}(0,P_{1})$, and $X_{2} \sim \mc{N}(0,P_{2})$.

Straightforward algebra  shows that the computation of \eqref{newup2} gives \eqref{equpbound}.

\vspace{0.5em}
\textit{\textbf{Lower Bound.}}
The proof of the lower bound follows by evaluating the equivocation rate in Theorem \ref{Th_inner} with a specific choice of the variables. More specifically, evaluating Theorem \ref{Th_inner} with the choice $C_{12}:=0$,  $U=V=\phi$, $V_1:=X_1$ and $V_2:=X_2$, with $X_1\sim\mc N(0,P_1)$ independent of $X_2\sim\mc N(0,P_2)$, and such that \eqref{condition} is satisfied, we obtain the rate expression in \eqref{eqbound}. The RHS of \eqref{eqbound} then follows by maximization over $\mathbb{E}[X_1^2] \le P_1$ and $\mathbb{E}[X_2^2] \le P_2$ and  satisfying \eqref{condition}.
\end{IEEEproof}
 
\vspace{0.5em}
\begin{remark}
The  bounds on the secrecy capacity in \eqref{equpbound} and \eqref{eqbound} have identical expressions but the maximization is over different sets of inputs. The bounds coincide in the case in which the inputs ($\mathbb{E}[X_1^2],\mathbb{E}[X_2^2]$) that maximize the RHS of \eqref{equpbound} also satisfy the condition \eqref{condition}. In this case, the perfect secrecy of the studied model is given  by
\begin{align}
\label{capacity}
C_s  =    \max\: \bigg[ \mc{C} \Big(\frac{|h_{1d}|^2\mathbb{E}[X_1^2] }{\sigma_{1}^2}\Big)  - \mc{C} \Big(\frac{|h_{1e}|^2\mathbb{E}[X_1^2] }{\sigma_{2}^2+|h_{2e}|^2\mathbb{E}[X_2^2] }\Big)\bigg]^+
\end{align}
where the maximization is over $\mathbb{E}[X_1^2] \le P_1$ and $\mathbb{E}[X_2^2] \le P_2$ satisfying
\begin{eqnarray}
  \mathcal{C}\Big(\frac{|h_{2d}|^2\mathbb{E}[X_2^2]}{|h_{1d}|^2\mathbb{E}[X_1^2]+{\sigma}_1^2}\Big) \ge   \mc{C} \Big(\frac{|h_{2e}|^2\mathbb{E}[X_2^2]}{\sigma_{2}^2}\Big).
\end{eqnarray}

\end{remark}
\vspace{0.5em}

\vspace{0.5em}
\subsubsection{Case $C_{12} := \infty $}
\vspace{0.5em}

As stated previously, in this case the model \eqref{gchan} reduces to a wiretap channel in which the transmitter equipped with two antenna and the legitimate receiver and eavesdropper equipped with single antennas. As it will be shown below, in this case the upper bound of Corollary~\ref{gaussian_upper_bound} and the lower bound of Corollary~\ref{cor1} coincide, thus providing a characterization of the secrecy capacity, which can also be obtained from \cite{oggier,Tie} in this specific case.
\vspace{0.5em}
\begin{corollary}
\label{cor4}
For the Gaussian model \eqref{gchan} with fully cooperating encoders, the secrecy capacity is given by
 \begin{eqnarray}
 \label{specialcase}
C_s =  \max_{\psi}[I(X_{1},X_{2};Y)-I(X_{1},X_{2};Z)]
 \end{eqnarray}
where  $[X_1,X_2] \sim \mc{N}(\bf{0},\bf{K}_{P})$ with ${{\mc{K}}_{P}}  = \Big \{ {\bf{K}_{P}} : {\bf{K}_{P}}$=  $\left [
 \begin{smallmatrix}
  P_{1} & \psi\sqrt{P_{1} P_{2}}\\
\psi\sqrt{P_{1} P_{2}} & P_{2}
 \end{smallmatrix} \right ]$, $-1\le \psi \le {1} \Big\}$, with $\mathbb{E}[X^2_{1}]$ and $\mathbb{E}[X^2_{2}]$ satisfying \eqref{p_con1}.
\normalsize
\end{corollary}
\vspace{0.5em}

\begin{IEEEproof}
The upper bound follows by Corollary~\ref{gaussian_upper_bound}. The proof of the lower bound follows by  evaluating the equivocation rate in Theorem~\ref{Th_inner} with a specific choice of the random variables. More specifically, the rate expression \eqref{specialcase} is obtained by setting $C_{12}:=\infty$, $U:=$ constant, $V_1:=X_1$, $ V=V_2=X_2$, in Theorem~\ref{Th_inner} where $[X_1,X_2] \sim \mc{N}(\bf{0},\bf{K}_{P})$ with ${{\mc{K}}_{P}}  = \Big \{ {\bf{K}_{P}} : {\bf{K}_{P}}$=  $\left [
 \begin{smallmatrix}
   P_{1} & \psi\sqrt{P_{1} P_{2}}\\
   \psi\sqrt{P_{1} P_{2}} & P_{2}
    \end{smallmatrix} \right ]$, $-1\le \psi \le {1} \Big\}$ and $\mathbb{E}[X^2_{1}]$ and $\mathbb{E}[X^2_{2}]$ satisfying \eqref{p_con1}.

\noindent With straightforward algebra, it can be checked that this corresponds also to the special case $C_{12}:=\infty$ in Corollary~\ref{cor1}.
\end{IEEEproof}

\section{Numerical results}
 \begin{figure}[t]
\psfragscanon
\begin{center}
\psfrag{User}[c][c][.75]{Encoder 1}
\psfrag{Destination}[c][c][.75]{Destination}
\psfrag{Eavesdropper}[c][c][.75]{Eavesdropper}
\psfrag{upper}[c][c][.75]{\hspace{4.5em}Upper Bound \eqref{guu}}
\psfrag{L211}[c][c][.75]{\hspace{9em}Lower Bound \eqref{lower_gaussian}, $C_{12}:=0$}
\psfrag{L222}[c][c][.75]{\hspace{9em}Lower Bound \eqref{lower_gaussian}, $C_{12}:=1$}
\psfrag{L224}[c][c][.75]{\hspace{9em}Lower Bound \eqref{lower_gaussian}, $C_{12}:=4$}
\psfrag{L226}[c][c][.75]{\hspace{9em}Lower Bound \eqref{lower_gaussian}, $C_{12}:=6$}
\psfrag{wiretap}[c][c][.75]{\hspace{3em} Wiretap Channel}
\psfrag{y}[c][c][.75]{Perfect Secrecy Rate (bits/channel use)}
\psfrag{x}[c][c][.75]{Location of Encoder 2}
\includegraphics[width=0.8\linewidth]{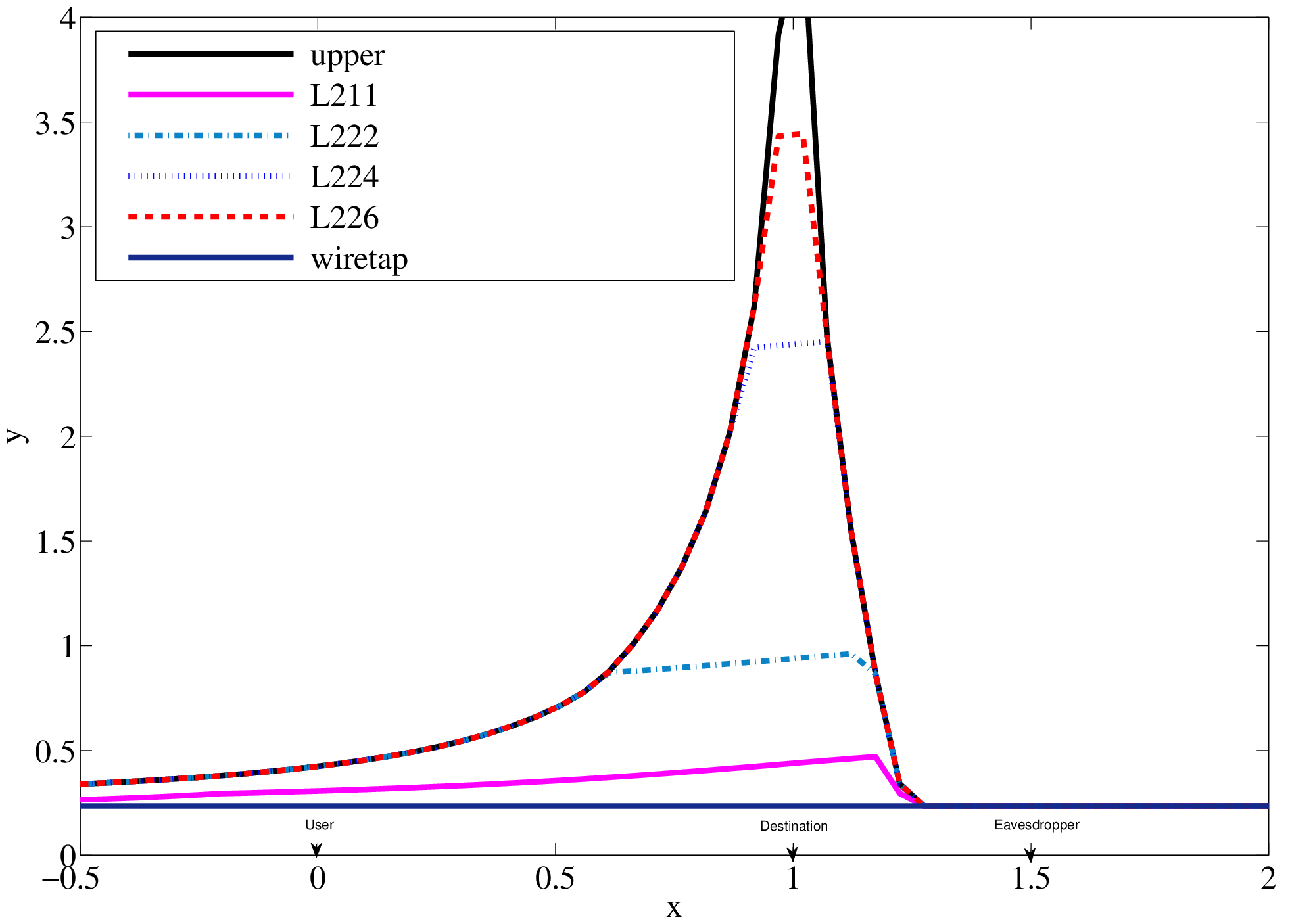}
\end{center}
\caption{Bounds on the secrecy capacity.}
\psfragscanoff
 \label{bounds}
\end{figure}

\begin{figure}[ht]
\psfragscanon
\begin{center}
\psfrag{L190}[c][c][.75]{\hspace{11.5em}Power used to inject noise, $C_{12}:=0$}
\psfrag{L191}[c][c][.75]{\hspace{11.5em}Power used to inject noise, $C_{12}:=1$}
\psfrag{L194}[c][c][.75]{\hspace{11.5em}Power used to inject noise, $C_{12}:=4$}
\psfrag{L196}[c][c][.75]{\hspace{11.5em}Power used to inject noise, $C_{12}:=6$}
\psfrag{L240}[c][c][.75]{\hspace{13em}Power used to trans. conf. inf., $C_{12}:=0$}
\psfrag{L241}[c][c][.75]{\hspace{13em}Power used to trans. conf. inf., $C_{12}:=1$}
\psfrag{L244}[c][c][.75]{\hspace{13em}Power used to trans. conf. inf., $C_{12}:=4$}
\psfrag{L246}[c][c][.75]{\hspace{13em}Power used to trans. conf. inf., $C_{12}:=6$}
\psfrag{y}[c][c][.75]{Power (watt)}
\psfrag{x}[c][c][.75]{Location of Encoder 2}
\includegraphics[width=0.8\linewidth]{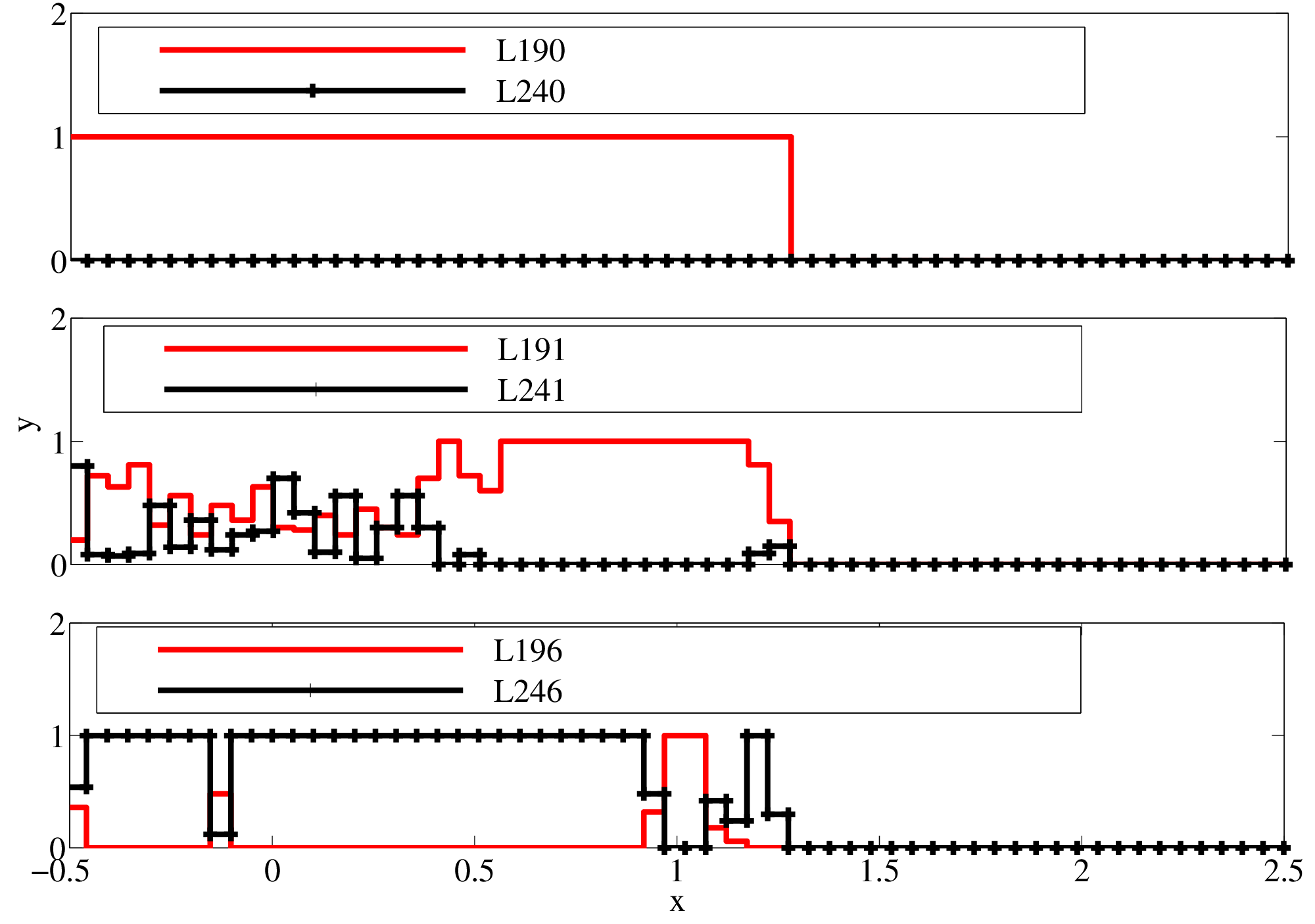}
\end{center}
\caption{Power splitting at Encoder 2 for different values of finite capacity link, where Encoder 1 is located at $(0,0)$,  the destination is located at $(1,0)$ and the eavesdropper is located at $(1.5,0)$.}
\psfragscanoff
 \label{PDF_power}
\end{figure}

\begin{figure}[ht]
\psfragscanon
\begin{center}
\psfrag{User}[c][c][.75]{Encoder 1}
\psfrag{dest}[c][c][.75]{\hspace{2em}Destination}
\psfrag{Eavesdropper}[c][c][.75]{Eavesdropper}
\psfrag{upper}[c][c][.75]{\hspace{4.5em}Upper Bound \eqref{guu}}
\psfrag{L211}[c][c][.75]{\hspace{9em}Lower Bound \eqref{lower_gaussian}, $C_{12}:=0$}
\psfrag{L222}[c][c][.75]{\hspace{9em}Lower Bound \eqref{lower_gaussian}, $C_{12}:=1$}
\psfrag{L224}[c][c][.75]{\hspace{9em}Lower Bound \eqref{lower_gaussian}, $C_{12}:=4$}
\psfrag{L226}[c][c][.75]{\hspace{9em}Lower Bound \eqref{lower_gaussian}, $C_{12}:=6$}
\psfrag{wiretap}[c][c][.75]{\hspace{3em} Wiretap Channel}
\psfrag{y}[c][c][.75]{Perfect Secrecy Rate (bits/channel use)}
\psfrag{x}[c][c][.75]{Location of Encoder 2}
\includegraphics[width=0.8\linewidth]{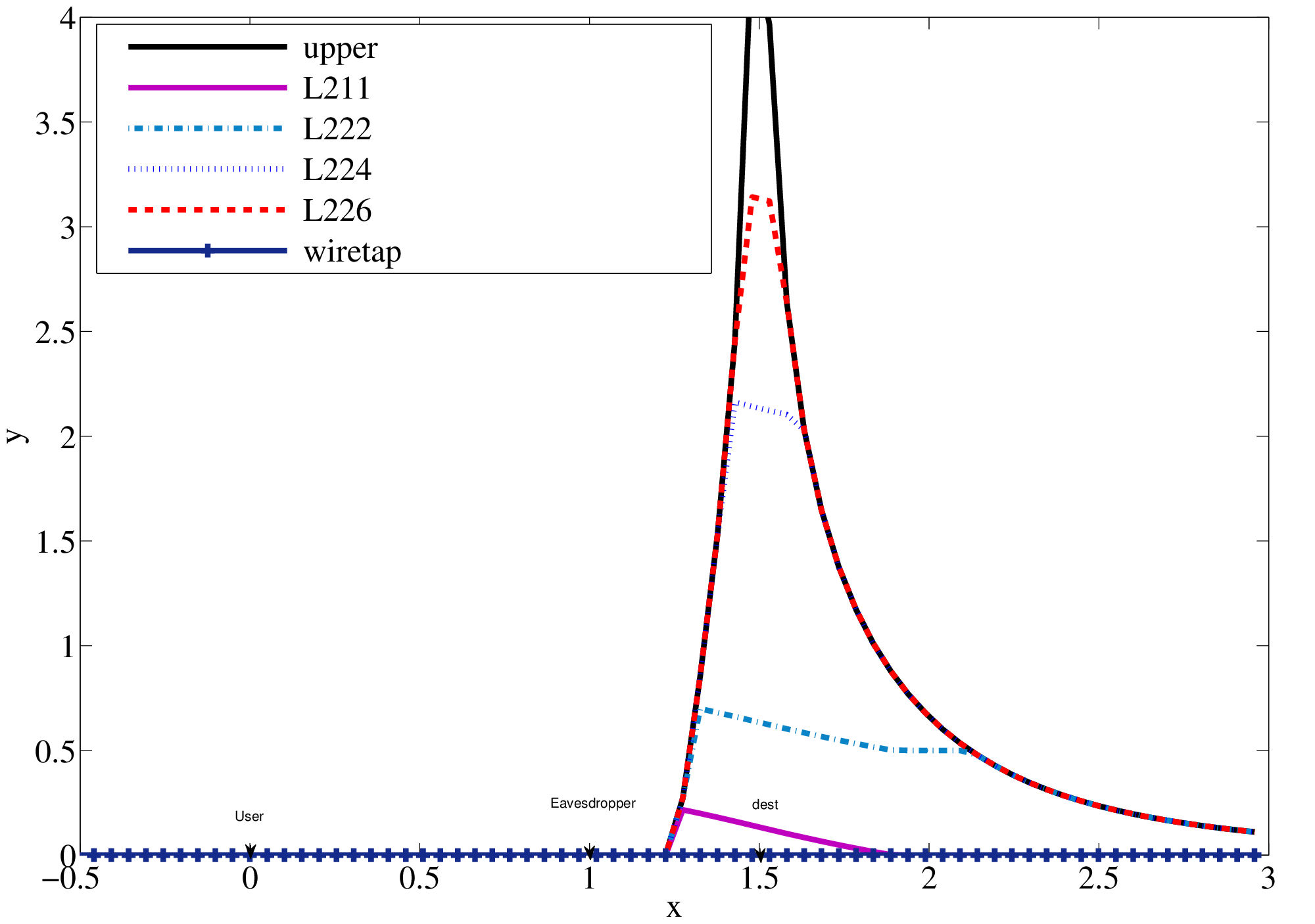}
\end{center}
\caption{Bounds on the secrecy capacity.}
\psfragscanoff
 \label{bounds_rev}
\end{figure}

\begin{figure}[ht]
\psfragscanon
\begin{center}
\psfrag{L190}[c][c][.75]{\hspace{11.5em}Power used to inject noise, $C_{12}:=0$}
\psfrag{L191}[c][c][.75]{\hspace{11.5em}Power used to inject noise, $C_{12}:=1$}
\psfrag{L194}[c][c][.75]{\hspace{11.5em}Power used to inject noise, $C_{12}:=4$}
\psfrag{L196}[c][c][.75]{\hspace{11.5em}Power used to inject noise, $C_{12}:=6$}
\psfrag{L240}[c][c][.75]{\hspace{13em}Power used to trans. conf. inf., $C_{12}:=0$}
\psfrag{L241}[c][c][.75]{\hspace{13em}Power used to trans. conf. inf., $C_{12}:=1$}
\psfrag{L244}[c][c][.75]{\hspace{13em}Power used to trans. conf. inf., $C_{12}:=4$}
\psfrag{L246}[c][c][.75]{\hspace{13em}Power used to trans. conf. inf., $C_{12}:=6$}
\psfrag{y}[c][c][.75]{Power (watt)}
\psfrag{x}[c][c][.75]{Location of Encoder 2}
\includegraphics[width=0.8\linewidth]{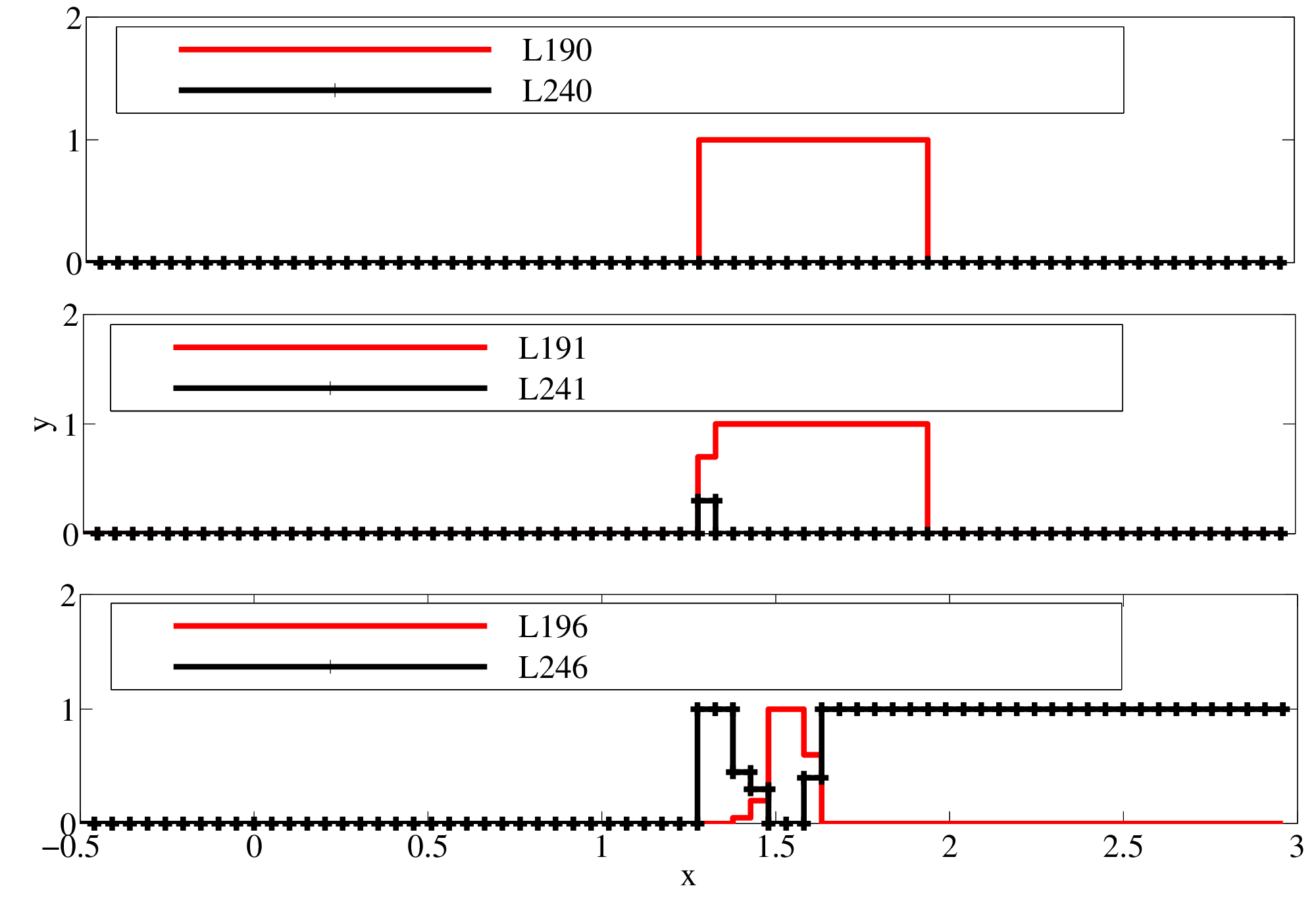}
\end{center}
\caption{Power splitting at Encoder 2 for different values of finite capacity link, where Encoder 1 is located at $(0,0)$, the destination is located at $(1.5,0)$ and the eavesdropper is located at $(1,0)$.}
\psfragscanoff
 \label{PDF_power_rev}
\end{figure}

In this section, we provide some numerical examples to illustrate our results. We consider the Gaussian MAC \eqref{gchan} in which the outputs at the legitimate receiver and eavesdropper are corrupted by additive white Gaussian noise (AWGN) of zero mean and unit variance each. We model channel gains between node $i \in \{1,2\}$ and $j \in \{d,e\}$ as distance dependent path loss, $h_{i,j} = {d_{i,j}^{-\gamma/2}}$, where $\gamma$ is the path loss exponent. We assume that both users have an average power constraint of 1 watt each and the path loss exponent $\gamma$:=2. We consider a network geometry in which Encoder 1 is located at the point $(0,0)$, Encoder 2 is located at the point $(d,0)$, the legitimate receiver is located at the point $(1,0)$ and the eavesdropper is located at the point $(1.5,0)$, where $d$ is the distance between Encoders 1 and 2. The upper  \eqref{guu} and the lower \eqref{lower_gaussian} bounds are optimized numerically for Gaussian inputs. Figure~\ref{bounds} shows the upper and lower bounds on the secrecy capacity for different values of finite capacity link. As a reference we consider the case in which there is no helping Encoder, i.e.,  a basic wiretap channel. If we set $C_{12}:=0$, Encoder 1 does not conference to Encoder 2,  for this setup the MAC \eqref{gchan} reduces to the classic WT-HI \cite{tang,xainghe}. In this case Encoder 2 can help Encoder 1 by injecting confusion codewords to confuse the eavesdropper \cite[Theorem 3]{lai}.

Figure~\ref{PDF_power} shows the power splitting  at Encoder 2 to transmit conferenced information and artificial noise in \eqref{lower_gaussian}, for different values of $C_{12}$. The region between $0.9 < d < 1.1$ is of particular interest where Encoder 2 is near to the destination. It can be easily seen that when helping encoder, Encoder 2, is near to the destination no power is allocated to transmit the conferenced information to the legitimate receiver and the lower bound is maximized by independent inputs. Roughly speaking, this follows because when Encoder 2 is near to the destination noise injection  provides higher secrecy rates.

If we increase the capacity of noiseless bit-pipe, the achievable secrecy rate increases, this follows because Encoder 2 is more informed about the information message from Encoder 1 and can cooperate with each other. For instance, if we consider a very large value of noiseless bit-pipe capacity, the upper and lower bounds will eventually coincide. This is due to the fact that a large value of $C_{12}$ results in full cooperation between the encoders, due to which the channel at hand reduces to a two-antenna transmitter wiretap channel for which secrecy capacity is established (Corollary \ref{cor4}).

Next, we consider a network geometry in which the eavesdropper is geographically placed at a more favorable location compared to the legitimate receiver. In this setting, compared to the earlier example we reverse the location of the destination and the eavesdropper, where the eavesdropper is located at the point $(1,0)$ and the legitimate receiver is located at the point $(1.5,0)$. Figure~\ref{bounds_rev} shows the optimized upper  \eqref{guu} and  lower \eqref{lower_gaussian} bounds for this case with different values of the finite capacity link. As a benchmark, similar to the previous example, we also plot the case in which there is no helper encoder (wiretap channel). From Figure~\ref{bounds_rev} it can be seen that in the absence of the helper encoder, it is not possible to obtain positive secrecy rates. This follows due to the fact that since the eavesdropper is located at a better position compared to the legitimate receiver, it can easily decode all the transmitted information. Roughly speaking, in this case degradedness condition is violated. With the presence of the helper Encoder, even though the  eavesdropper is at a more favorable position compared to the legitimate receiver, one can still obtain positive secrecy rates. This follows because, in this setting Encoder 2 can help the legitimate receiver by injecting statistically independent artificial noise to confuse the eavesdropper which in turns provides higher secrecy rates~\cite[Theorem 3]{lai}. From Figure~\ref{bounds_rev}, one can also see that for large values of $C_{12}$, the secrecy rate increases. This follows, since  Encoder 2 is more informed about the confidential messages at the Encoder 1, cooperation between Encoders can provides higher secrecy rates (Corollary \ref{cor4}). Figure~\ref{PDF_power_rev} shows the power splitting at Encoder 2 to transmit conferenced information and artificial noise for the new setup. In this setting, when the Encoder 2 is located between the eavesdropper and the legitimate receiver, it uses full power to inject artificial noise to confuse the eavesdropper which provides higher rates.

\section{Conclusion}
We studied a special case of Willems's multi-access channel with partially cooperating encoders \cite{willems} in the presence of a passive eavesdropper, from a security viewpoint. For the general DM case, we established outer and inner bounds on the capacity-equivocation region. The inner bound is obtained by a combination of Willems's coding scheme, injection of statistically independent artificial noise \cite[Theorem 3]{lai} and binning for security. The outer bound is obtained by extending outer bounding techniques that are developed previously in the context of broadcast channels with confidential messages and Willems's model to the studied model. The developed outer and inner bounds do not agree in general, but can be seen as a step ahead towards characterizing the capacity-equivocation region. For the Gaussian setup, we focus on the case of perfectly secure transmission, and establish lower and upper bounds on the secrecy capacity. We also study some extreme cases of cooperation between the encoders. For the case in which the encoders do not cooperate the considered setup reduces to wiretap-channel with an external helper interferer, a setup whose secrecy capacity is still unknown \cite{tang,xainghe,LYT08}. For this particular setup, we show that under certain conditions our lower and upper bounds coincide, and so we characterize the secrecy capacity fully. For the case of full cooperation between the encoders, the studied setup reduces to a wiretap channel in which the transmitter is equipped with two-antenna, and the legitimate receiver and eavesdropper are equipped with single antennas. In this case,  the developed bounds agree, and so we obtain the secrecy capacity expression.

\appendices
\section{Proof of Theorem 1}
\label{appendix_outer}
The converse uses elements from the proof given in the context of broadcast channels with confidential messages \cite{csiszar} and the proof established in the context of multiple access channel with partially cooperating encoders \cite{willems}. We begin the proof by first setting $W_{12}:=G_1^K$. \\
\vspace{.5em}\\
\noindent\textit{\textbf{Bounds on the equivocation rate.}}
\vspace{.5em}\\
\textbf{a)} We first bound the equivocation rate $R_e$ as follows.
\setlength{\arraycolsep}{0.2em}

\begin{eqnarray}
nR_{e}&=& {H}(W | Z^n )\notag \\
&\overset{(a)}{=}& {H}(W, W_{12}| Z^n ) \notag\\
&=& H(W, W_{12})- I(W, W_{12}; Z^n)\notag\\
&=& I(W, W_{12}; Y^n)+{H}(W, W_{12} | Y^n)- I(W,W_{12}; Z^n)\notag\\
&\overset{(b)}{=}& I(W, W_{12}; Y^n)+{H}(W| Y^n)- I(W,W_{12}; Z^n)\notag\\
&\overset{(c)}{\le}& \sum_{i=1}^n I(W,W_{12};Y_i| Y^{i-1})-I(W,W_{12} ;Z_i | Z_{i+1}^n)+n\epsilon_{n} \notag\\
&=&\sum_{i=1}^n  I(W,W_{12}, Z_{i+1}^n; Y_i |  Y^{i-1} )   - I(Z_{i+1}^n; Y_i  | W,W_{12},  Y^{i-1})
- I(W,W_{12}, Y^{i-1} ; Z_i |  Z_{i+1}^n )  \notag\\ &&+ I(Y^{i-1};Z_i | W,W_{12},  Z_{i+1}^n )+n\epsilon_{n}\notag\\
&\overset{(d)}{=}& \sum_{i=1}^n I(W,W_{12}, Z_{i+1}^n; Y_i |  Y^{i-1} )  - I(W,W_{12}, Y^{i-1} ; Z_i |  Z_{i+1}^n )+n\epsilon_{n} \notag\\
&=& \sum_{i=1}^n  I(Z_{i+1}^n; Y_i |  Y^{i-1} )+  I(W,W_{12} ; Y_i |  Y^{i-1}, Z_{i+1}^n)-I(Y^{i-1} ; Z_i |  Z_{i+1}^n )\notag\\&&-I(W,W_{12}  ; Z_i |  Y^{i-1}, Z_{i+1}^n )+n\epsilon_{n}\notag\\
\label{proof1b}
&\overset{(e)}{=}& \sum_{i=1}^n  I(W,W_{12} ; Y_i |  Y^{i-1}, Z_{i+1}^n) -I(W,W_{12} ; Z_i |  Y^{i-1}, Z_{i+1}^n )+n\epsilon_{n}
\end{eqnarray}
where $\epsilon_n \rightarrow 0$ as $n \rightarrow \infty$;
 $(a)$ and $(b)$ follow because $W_{12}$ is a function of $W$, $(c)$ follows from Fano's inequality; and $(d)$ and $(e)$ follows from Lemma 7 in \cite{csiszar}.
 
Let us define $\bar{U}_i :=  Y^{i-1}, Z_{i+1}^n$,   $\bar{V}_{1,i} :=    W, Z_{i+1}^n$, and $\bar{V}_{2,i} :=  W_{12}, Y^{i-1}$. We introduce a random variable $T$ uniformly distributed over $\{1,2,\cdots,n\}$ and define $\bar{U}:=\bar{U}_T$,   $\bar{V}_1 :=\bar{V}_{1,T}$, $\bar{V}_2 :=\bar{V}_{2,T}$, $X_{1}:=X_{1,T}$, $X_{2}:=X_{2,T}$, $Y:=Y_{T}$, $Z:=Z_{T}$. Also, we let $U := (T,\bar{U})$,   $V_{1} := (T,\bar{V}_{1})$, $V_{2} := (T,\bar{V}_{2})$.

\vspace{0.2cm}

\noindent Thus, we have
\vspace{-0.6cm}

\begin{eqnarray}
R_{e}&\le& \frac{1}{n}\sum_{i=1}^n  I(W,W_{12} ; Y_i |  Y^{i-1},Z_{i+1}^n)  -I(W,W_{12} ; Z_i |  Y^{i-1}, Z_{i+1}^n )+n\epsilon_{n}\notag \\
&=& \frac{1}{n}\sum_{i=1}^n  I(W,W_{12},Y^{i-1}, Z_{i+1}^n ; Y_i |  Y^{i-1}, Z_{i+1}^n)  -I(W,W_{12},  Y^{i-1}, Z_{i+1}^n ; Z_i |  Y^{i-1}, Z_{i+1}^n )+\epsilon_{n}\notag \\
&\overset{(f)}{=}&  \frac{1}{n} \sum_{i=1}^n I(\bar{V}_{1,i},\bar{V}_{2,i} ;Y_{i}| \bar{U}_{i})-I(\bar{V}_{1,i},\bar{V}_{2,i} ;Z_{i}| \bar{U}_{i})+\epsilon_{n}\notag\\
&{=}&  I(\bar{V}_{1},\bar{V}_{2} ;Y| \bar{U},T)-I(\bar{V}_{1},\bar{V}_{2} ;Z| \bar{U},T)+\epsilon_{n}\notag\\
&{=}&  I(\bar{V}_{1},\bar{V}_{2},T ;Y| \bar{U},T)-I(\bar{V}_{1},\bar{V}_{2}, T;Z| \bar{U},T)+\epsilon_{n}\notag\\
&\overset{(g)}{=}& I(V_{1},V_{2} ;Y| U)-I(V_{1},V_{2} ;Z| U)+\epsilon_{n}
\end{eqnarray}
where $(f)$ follows from the definition of $\bar{U}_i,\bar{V}_{1,i}$ and $\bar{V}_{2,i}$; and $(g)$ follows from the definition of ${U},{V}_{1}$ and ${V}_{2}$.
\vspace{.5em}\\
 
We can also bound the equivocation rate $R_e$ as follows. We continue from \eqref{proof1b} to get
\begin{eqnarray}
\label{proof2}
R_e&{\leq}& \frac{1}{n}\sum_{i=1}^n  I(W,W_{12} ; Y_i |  Y^{i-1}, Z_{i+1}^n)  -I(W, W_{12} ; Z_i |  Y^{i-1}, Z_{i+1}^n )+\epsilon_{n}\notag\\
&{=}& \frac{1}{n}\sum_{i=1}^n  I( W; Y_i | W_{12}, Y^{i-1}, Z_{i+1}^n) +I(W_{12} ; Y_i |  Y^{i-1}, Z_{i+1}^n)  -I(W, W_{12} ; Z_i |  Y^{i-1}, Z_{i+1}^n )+\epsilon_{n}\notag\\
&\overset{(h)}{\leq}& \frac{1}{n}\sum_{i=1}^n  I(W ; Y_i | W_{12}, Y^{i-1}, Z_{i+1}^n) +H(W_{12})  -I(W, W_{12} ; Z_i |  Y^{i-1}, Z_{i+1}^n )+\epsilon_{n}\notag\\
&\overset{(i)}{\le}& \frac{1}{n}\sum_{i=1}^n  I(W ; Y_i | W_{12}, Y^{i-1}, Z_{i+1}^n)   -I(W, W_{12} ; Z_i |  Y^{i-1}, Z_{i+1}^n )+C_{12}+\epsilon_{n}\notag\\
&{=}& \frac{1}{n} \sum_{i=1}^n  I(W,  Z_{i+1}^n; Y_i | W_{12},  Y^{i-1}, Z_{i+1}^n)   -I(W, W_{12}, Y^{i-1}, Z_{i+1}^n ; Z_i |  Y^{i-1}, Z_{i+1}^n )+C_{12}+\epsilon_{n}\notag\\
&\overset{(j)}{=}& \frac{1}{n} \sum_{i=1}^n I(\bar{V}_{1,i} ;Y_{i}| \bar{V}_{2,i}, \bar{U}_{i})   -I(\bar{V}_{1,i},\bar{V}_{2,i} ;Z_{i}| \bar{U}_{i})+C_{12} +\epsilon_{n}\notag\\
&{=}&  I(\bar{V}_{1} ;Y| \bar{V}_{2}, \bar{U},T)  -I(\bar{V}_{1},\bar{V}_{2} ;Z| \bar{U},T)+C_{12}+\epsilon_{n}\notag\\
&{=}&  I(\bar{V}_{1},T ;Y| \bar{V}_{2}, \bar{U},T)  -I(\bar{V}_{1},\bar{V}_{2}, T;Z| \bar{U},T)+C_{12}  +\epsilon_{n}\notag\\
&\overset{(k)}{=}& I(V_{1};Y| {V}_{2}, U)  -I(V_{1},V_{2} ;Z| U)+C_{12}+\epsilon_{n}
\end{eqnarray}
where  $(h)$ follows because \( I(W_{12};Y_i |  Y^{i-1}, Z_i ^n) \le H(W_{12}|Y^{i-1}, Z_i ^n) \le H(W_{12})\), $(i)$ follows because $H(W_{12}) \le \sum_{k=1}^K H(G_{1k}) \le \sum_{k=1}^K \log(|\mc {G}_{1k}|) \le n C_{12}$,    $(j)$ follows from the definition of $\bar{U}_i, \bar{V}_{1,i}$ and $\bar{V}_{2,i}$; and $(k)$ follows from the definition of ${U}, {V}_{1}$ and ${V}_{2}$.
 
\vspace{.5em}

\noindent\textit{\textbf{Bounds on the transmission rate.}}\vspace{.5em}\\
\textbf{b)} We now bound the transmission rate $R$ as follows.
\begin{eqnarray}
\label{achrate}
nR&=& {H}(W)\notag \\
&\overset{(l)}{=}& {H}(W,W_{12})\notag \\
&=& I(W,W_{12};Y^n)+H(W,W_{12} | Y^n)\notag\\
&\overset{(m)}{=}& I(W,W_{12};Y^n)+H(W | Y^n)\notag\\
&\overset{(n)}{\le}&  I(W,W_{12};Y^n)+ n \epsilon_{n}\notag \\
&=&  \sum_{i=1}^n I(W,W_{12};Y_{i}| Y^{i-1})+ n \epsilon_{n} \notag\\
&=&  \sum_{i=1}^n H(Y_{i}| Y^{i-1})-H(Y_{i}| W,W_{12}, Y^{i-1})+ n \epsilon_{n} \notag\\
&\overset{(o)}{\le}&  \sum_{i=1}^n H(Y_{i})-H(Y_{i}| W,W_{12}, Y^{i-1})+ n \epsilon_{n} \notag\\
&\overset{(p)}{\le}&  \sum_{i=1}^n H(Y_{i})-H(Y_{i}| W,W_{12},  Y^{i-1}, Z_{i+1}^n)+ n \epsilon_{n} \notag\\
&=&  \sum_{i=1}^n I(W,W_{12},  Y^{i-1}, Z_{i+1}^n;Y_{i}) + n \epsilon_{n}
\end{eqnarray}
where $(l)$ and $(m)$ follow because $W_{12}$ is a function of $W$, $(n)$ follows from Fano's inequality; $(o)$ and $(p)$ follow from the fact that conditioning reduces entropy.

We continue from \eqref{achrate} to get
\begin{eqnarray}
R&\le&  \frac{1}{n} \sum_{i=1}^n I(W,W_{12},  Y^{i-1}, Z_{i+1}^n;Y_{i}) +  \epsilon_{n} \notag\\
&\overset{(q)}{=}&  \frac{1}{n} \sum_{i=1}^n I(\bar{V}_{1,i},\bar{V}_{2,i} ;Y_{i})+\epsilon_{n}\notag\\
&{=}&  I(\bar{V}_{1},\bar{V}_{2} ;Y| T)+\epsilon_{n}\notag\\
&{=}&  I(\bar{V}_{1},\bar{V}_{2},T ;Y)-I(T;Y)+\epsilon_{n}\notag\\
&\le&  I(\bar{V}_{1},\bar{V}_{2}, T;Y)+\epsilon_{n}\notag\\
&\overset{(r)}{=}& I(V_{1},V_{2} ;Y)+\epsilon_{n}
\end{eqnarray}
where $(q)$ follows from the definition of $\bar{V}_{1,i}$ and $\bar{V}_{2,i}$; and $(r)$ follows from the definition of ${V}_{1}$ and ${V}_{2}$.

\vspace{.5em}
We can also bound the transmission $R$ as follows
\begin{eqnarray}
\label{achrate2}
nR&=& {H}(W)\notag \\
&\overset{(s)}{=}& {H}(W,W_{12})\notag \\
&=& {H}(W|W_{12})+{H}(W_{12})\notag \\
&\overset{(t)}{\le}& {H}(W|W_{12})+nC_{12}\notag \\
&=& I(W; Y^n|W_{12})+H(W|W_{12}, Y^n)+nC_{12}\notag\\
&\overset{(u)}{\le}& I(W; Y^n|W_{12})+H(W|Y^n)+nC_{12}\notag\\
&\overset{(v)}{\le}&  \sum_{i=1}^n I(W;Y_{i}| W_{12},Y^{i-1})+ nC_{12}+ n\epsilon_{n} \notag\\
&=&  \sum_{i=1}^n H(Y_{i}| W_{12},Y^{i-1})-H(Y_{i}|W,W_{12}, Y^{i-1}) + nC_{12}+n\epsilon_{n} \notag\\
&\overset{(w)}{\le}&  \sum_{i=1}^n H(Y_{i}|W_{12},Y^{i-1})-H(Y_{i}|  W,W_{12}, Y^{i-1}, Z_{i+1}^n )  + nC_{12}+ n\epsilon_{n} \notag\\
&{=}& \sum_{i=1}^n I(W, Z_{i+1}^n;Y_{i}| W_{12}, Y^{i-1}) + nC_{12}+n\epsilon_{n}
\end{eqnarray}
where $(s)$ follows because $W_{12}$ is a function of $W$, $(t)$ follows because $H(W_{12}) \le \sum_{k=1}^K \notag H(G_{1k}) \le \sum_{k=1}^K \log(|\mc {G}_{1k}|) \le n C_{12}$, $(u)$ and $(w)$ follow from the fact that conditioning reduces entropy; and $(v)$ follows from Fano's inequality.

We continue from \eqref{achrate2} to get
\begin{eqnarray}
R&\le&  \frac{1}{n} \sum_{i=1}^n I(W,  Z_{i+1}^n;Y_{i}| W_{12}, Y^{i-1}) +  C_{12} +\epsilon_{n} \notag\\
&\overset{(x)}{=}&  \frac{1}{n} \sum_{i=1}^n I(\bar{V}_{1,i} ;Y_{i}|\bar{V}_{2,i})+  C_{12}+ \epsilon_{n}\notag\\
&{=}&  I(\bar{V}_{1} ;Y| \bar{V}_{2}, T)+C_{12}+\epsilon_{n}\notag\\
&=&  I(\bar{V}_{1}, T;Y| \bar{V}_{2}, T)+C_{12}+\epsilon_{n}\notag\\
&\overset{(y)}{=}& I(V_{1} ;Y| V_{2} )+C_{12}+\epsilon_{n}
\end{eqnarray}
where $(x)$ follows from the definition of $\bar{V}_{1,i}$ and $\bar{V}_{2,i}$; and $(y)$ follows from the definition of   ${V}_{1}$ and ${V}_{2}$.

This completes the proof of Theorem 1.
 
\section{Proof of Theorem 2}
\label{appendix_inner}
\begin{figure}
\psfragscanon
\begin{center}
\psfrag{u}[l][][1.5]{\hspace{ 0.75em}$U$ }
\psfrag{v}[l][][1.5]{\hspace{-.15em}$V_1$ }
\psfrag{x}[l][][1.5]{\hspace{-.4em}$X_1$ }
\psfrag{w}[l][][1.5]{\hspace{-.25em}$V $ }
\psfrag{y}[l][][1.5]{\hspace{-.2em}$V_2$ }
\psfrag{z}[l][][1.5]{\hspace{-.4em}$X_2$ }
\includegraphics[width=\linewidth]{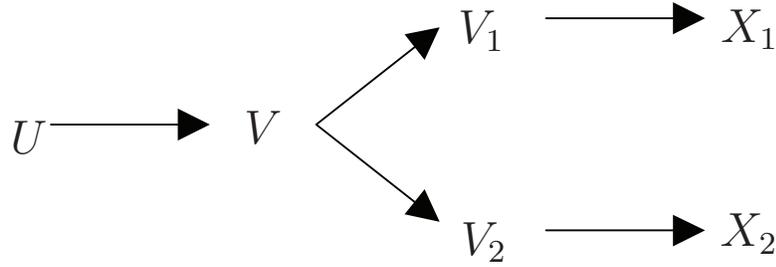}
\end{center}
\caption{Relationship between auxiliary random variables.}
\psfragscanoff
 \label{auxilary}
\end{figure}

The proof is a combination of  Willems's coding scheme \cite{willems} and noise forwarding scheme established by Lai \textit{et al.} \cite{lai} with additional binning for security \cite{wyner}. We begin the proof by first setting $V_1:=X_1$, $V_2:=X_2$ in Theorem 2. After proving Theorem 2 with $X_1,X_2$, we prefix a memoryless channel $p(x_1,x_2|v_1,v_2)=p(x_1 |v_1 )p( x_2| v_2) $ as reasoned in \cite[Lemma 4]{csiszar} to finish the proof. Figure~\ref{auxilary} shows the relationship between the auxiliary random variables.
\vspace{.5em}\\
\noindent\textit{\textbf{Random Coding.}}
\begin{enumerate}
\item Randomly generate a typical sequence $u^n$ with probability $p(u^n)=\prod_{i=1}^n p(u_i)$. We assume that all terminals know $u^n$.

\item For each $u^n$ randomly generate $2^{nR_{12}}$ independent and identically distributed (i.i.d) $v^n$ codewords, each with probability $p(v^n|u^n)=\prod_{i=1}^n p(v_{i}|u_i)$ and index them as $v^n(w_0)$, $w_0 \in  [1,2^{n R_{12}}]$, where we set $R_{12} \le C_{12}$.

\item For each $v^n(w_0)$ generate $2^{nR_1}$ conditionally i.i.d $x_1^n$ sequence, each with probability $p(x_1^n|v^n(w_0), u^n)=\prod_{i=1}^n p (x_{1i}|v_{i}(w_0),u_i)$, and index them as $x_1^n(w_1,w_0)$, $w_1 \in  [1,2^{n R_1}]$.

\item For each $v^n(w_0)$ generate $2^{nR_2}$ conditionally i.i.d $x_2^n$ sequence, each with probability $p(x_2^n|v^n( w_0),   u^n)=\prod_{i=1}^n p (x_{2i}|v_{i}( w_0) ,u_i)$, and index them as $x_2^n(w_c, w_0)$, $w_c \in [1,2^{n R_2}]$, where we set $R_2 = \min\{I(X_2;Y|V,  U),I(X_2;Z|X_1,V,  U)\}-\epsilon $.

\end{enumerate}
 
We define
 
\begin{equation}
\label{final_eq}
R''=R'+ \min\{I(X_2;Y|V, U),I(X_2;Z|X_1,V, U)\} -I(X_1,X_2;Z|U),
\end{equation}

\noindent and $R'=R_1+R_{12}$,  where $\mc{W''} = \{1,\hdots, 2^{nR''} \}, \mc{L} = \{1,\hdots, 2^{n(R'-R'')}\}$ and $\mc{K} = \mc{W''}\times \mc{L}$. In the following we assume that $R'' \ge 0$, otherwise this coding scheme does not achieve any security level.
\vspace{.5em}\\
\noindent\textit{\textbf{Encoding.}} \\
For a given rate-equivocation pair $(R,R_e)$ with $R \le R'$ and $R_e \le R$, we propose the following random coding scheme. Let $w \in \mc{W}= \{1,\hdots 2^{nR}\}$ be the total number of message. The stochastic encoder performs the mapping as follows.
\begin{itemize}
\item If $R \ge R''$, then  let $\mc{W} = \mc{W''}\times \mc{J}$ where $\mc{J} = \{1,\hdots, 2^{n(R-R'')}\}$. Let $g$ be the mapping that partitions $\mc {L}$ into $\mc {J}$ subsets of nearly equal size. The stochastic encoder then maps $w = (w'',j) \rightarrow (w'',l)$, where $l$ is uniformly chosen from $g^{-1}(j) \subset \mc {L}$. We define $t=(w'',l)$, where $t$ is further partition into $w_1$ and $w_0$ of rates $R_1$, and  $R_{12}$ respectively.

\item If $R \le R''$, the stochastic encoder maps  $w \rightarrow  (w,l)$, where $l$ is uniformly chosen from $\mc{L}$. Next we define $t=(w,l)$, where $t$ is further partition into $w_1$ and $w_0$ of rates $R_1$, and  $R_{12}$ respectively.
\end{itemize}
After the mapping, Encoder 1 transmits $x_1^n(w_1,w_0)$ and Encoder  2 transmits $x_2^n(w_c, w_0)$, where Encoder 2 randomly selects $w_c \in \{1,\hdots, 2^{nR_2} \}$.
\vspace{.5em}\\
\noindent \textit{\textbf{Decoding.}} \\
The legitimate receiver performs the decoding as follows.
\begin{itemize}
\item After the conferencing process, Encoder 2  knows $w_0$, if $R_{12} \le C_{12}$.

\item The legitimate receiver declares that $(\hat{w}_0=w_0)$ was sent, by looking at jointly $\epsilon$-typical $(x_2^n(w_c,w_0),y^n, u^n)$.

\item The legitimate receiver then declares that $(\hat{w}_c=w_c)$ was sent, by looking at jointly $\epsilon$-typical $(x^n_2(w_c,\hat{w}_0),y^n, u^n)$.

\item Afterwards the legitimate receiver declares that $(\hat{w}_1=w_1)$, if  $(x_1^n({w}_1,\hat{w}_0),x_2^n(\hat{w}_c, \hat{w}_0),y^n,u^n)$ is jointly $\epsilon$-typical.
\end{itemize}
\vspace{.5em}
\noindent \textit{\textbf{Probability of Error Analysis.}} \\
To transmit $(w_1,w_0)$ to the legitimate receiver, Encoder 1  and Encoder 2 transmit $x_1^n(w_1,w_0)$ and $x_2^n(w_c, w_0)$ respectively. Due to the symmetry of random code construction, the average error probability does not depends on the particular message index that was sent. Thus without loss of generality we consider that $(w_1,w_0)=(1,1)$ was sent and
define the error events
\begin{eqnarray*}
E_{w_1w_0} = \{( v^n( w_0),x_1^n(w_1,w_0),x_2^n (w_c, w_0), y^n | u^n) \in T_\epsilon^{n}\} .
\end{eqnarray*}
The error occurs if the transmitted and received codewords are not jointly typical $(E_{11}^c)$ or when a wrong codeword is jointly typical with the received codewords $(E_{w_11}$ or $E_{w_1w_0})$.
The probability of decoding an error is given by
\begin{eqnarray}
\label{error}
P_e^n \le P(E_{11}^c)+\sum_{\substack{w_1\ne 1,\\w_0=1}} P(E_{w_11})+\sum_{\substack{w_1\ne 1,\\w_0\ne1}} P(E_{w_1w_0}).
\end{eqnarray}
The first term, $P(E_{11}^c) \rightarrow 0$ by AEP \cite[Chapter 3]{cover}.
Now we consider the second term in \eqref{error} as follows


\begin{eqnarray}
\label{neweq}
 \sum_{\substack{w_1\ne 1,\\w_0=1}}  P(E_{w_1 1}) &\le& 2^{nR_1}\sum_{\substack{( v^n,x_1^n,x_2^n,\\y^n|u^n) \in T_\epsilon^{n}}}  p(v^n| u^n)p(x_1^n|v^n, u^n) p(x_2^n|v^n,u^n)p(y^n|x_2^n,v^n,  u^n) \notag\\
&\overset{(a)}{\le}&  2^{nR_1} 2^{n(H( V,X_1,X_2,Y|U)+\epsilon)}   2^{-n(H(V|U)-\epsilon)} 2^{-n(H(X_1|V,U)-2\epsilon)}2^{-n(H(X_2|V, U)-2\epsilon)}\notag\\&&.2^{-n(H(Y|X_2,V,U)-2\epsilon)} \notag\\
&=&  2^{nR_1} 2^{-n[ H(V|U)+H(X_1|V,U)+H(X_2|V, U)- 5\epsilon]} 2^{-n[H(Y|X_2,V, U)-H( V,X_1,X_2,Y|U)-3\epsilon]}   \notag
\end{eqnarray}



\noindent
where $(a)$ follows from the joint AEP \cite[chapter 14]{cover}.

\noindent Thus if
\setlength{\arraycolsep}{0.1em}


\begin{eqnarray}
\label{rate12}
R_1 &\le&  H(V| U)+H(X_1|V,U) +H(X_2|V, U)   +H(Y|X_2,V, U)-H( V,X_1,X_2,Y|U)-8\epsilon \notag\\
           &=& I (X_1;Y|X_2,V, U)-8\epsilon
\end{eqnarray}



\noindent the second term in \eqref{error} goes to zero as $n\rightarrow \infty$. From random code construction it follows that
\setlength{\arraycolsep}{0.2em}
$$R_1+R_{12} \le I (X_1;Y|X_2,V,U)+C_{12}-8\epsilon. $$

\noindent Next, we consider the third term in \eqref{error} as follows


\begin{eqnarray}
 \sum_{\substack{w_1\ne 1,\\w_0\ne1}} P(E_{w_1 w_0}) &\le
& 2^{n(R_1+R_{12})}  \sum_{\substack{(v^n,x_1^n,x_2^n,\\y^n|u^n) \in T_\epsilon^{n}}}  p(v^n| u^n)p(x_1^n|v^n, u^n) p(x_2^n|v^n,u^n)p(y^n|u^n)\notag\\
&\overset{(b)}{\le}&  2^{n(R_1+R_{12})} 2^{n(H( V,X_1,X_2,Y|U)+\epsilon)} 2^{-n(H(V| U)- \epsilon)} 2^{-n(H(X_1|V,U)-2\epsilon)} 2^{-n(H(X_2|V, U)-2\epsilon)}\notag\\&&.2^{-n(H(Y|U)-\epsilon)}\notag\\
&=& 2^{n(R_1+R_{12})}  2^{-n[ H(V| U)+H(X_1|V,U)+H(X_2|V, U)-5\epsilon]}  2^{-n[H(Y|U)-H( V,X_1,X_2,Y|U)-2\epsilon]}\notag
\end{eqnarray}



\noindent where $(b)$ follows from the joint AEP \cite[chapter 14]{cover}.

\noindent Thus if


\begin{eqnarray}
R_1+R_{12} &\le&  H(V| U)+H(X_1|V,U)  +H(X_2|V, U)  +H(Y|U) -H( V,X_1,X_2,Y|U)-7\epsilon \notag\\
           &=& I (X_1,X_2,V;Y|U)-7\epsilon\notag\\
           &=& I (X_1,X_2 ;Y|U)-7\epsilon
\end{eqnarray}


\noindent the third term in \eqref{error}  goes to zero as $n\rightarrow \infty$.

Therefore for a sufficiently large values of $n$, the probability of error goes to zero, if
\begin{eqnarray}
R' \le \min\{I(X_1;Y | X_2,V,  U)+C_{12},  I(X_1, X_2 ;Y | U) \}.\notag
\end{eqnarray}

\vspace{.5em}
\noindent\textit{\textbf{Equivocation computation.}}\\
The computation of equivocation is given as follows.
\setlength{\arraycolsep}{0.1em}

\begin{eqnarray}
H(W|Z^n)&\ge& H(W|Z^n,U^n) \notag\\
                &=& H(W,Z^n|U^n)-H(Z^n|U^n) \notag\\
                &=& H(W,V^n,X_1^n,X_2^n,Z^n|U^n) -H(V^n,X_1^n,X_2^n|W,Z^n,U^n)-H(Z^n|U^n) \notag\\
                &=& H(V^n,X_1^n,X_2^n|U^n) +H(W,Z^n|V^n,X_1^n,X_2^n,U^n) -H(V^n,X_1^n,X_2^n|W,Z^n,U^n)\notag\\&&-H(Z^n|U^n) \notag\\
                &\ge& H(V^n,X_1^n,X_2^n|U^n)+H(Z^n|V^n, X_1^n,X_2^n,U^n)  -H(V^n,X_1^n,X_2^n|W,Z^n,U^n)\notag\\&&-H(Z^n|U^n).
\end{eqnarray}


\noindent We first consider $H(V^n,X_1^n,X_2^n|W,Z^n,U^n)$. Given $W$ the eavesdropper only needs to decode $l$, and $w_c$, which can be decoded because


\begin{eqnarray}
\label{eq_L}
\frac{1}{n}\log(|R'-R''|)+R_2 &=& R'-R'+I(X_1,X_2;Z|U) - \min\{I(X_2;Y|V, U),I(X_2;Z|X_1,V, U)\}\notag\\&&+\min\{I(X_2;Y|V,U),I(X_2;Z|X_1,V,U)\}-\epsilon\notag\\
&\le&  I(X_1,X_2;Z|U). \notag
\end{eqnarray}



\noindent Therefore, it can be easily shown that,
\begin{eqnarray}
H(V^n,X_1^n,X_2^n|W,Z^n,U^n) \le \epsilon_2.
\end{eqnarray}
\noindent Since the channel is memoryless we can write

\begin{eqnarray}
H(Z^n|U^n)-H(Z^n|V^n,X_1^n,X_2^n,U^n) &\le& nI(X_1,X_2,V ;Z|U)+n\epsilon_n \notag\\
        &=& nI(X_1,X_2;Z|U)+n\epsilon_n
\end{eqnarray}

 

\noindent where $\epsilon_n\rightarrow 0$, as  $n\rightarrow \infty$ \cite{wyner}.
If  $R \ge R''$ then $H(V^n,X_1^n,X_2^n|U^n)= H(V^n,X_1^n |U^n)+H(X_2^n|V^n,X_1^n,U^n) =   nR'+nR_2 $, which follows from codebook construction.
The secrecy rate is then given by
\begin{eqnarray}
nR_e &\ge& n(R'+R_2-I(X_1,X_2;Z|U)-\epsilon_3).
\end{eqnarray}
If $R \le R''$,  $H(V^n,X_1^n,X_2^n|U^n) = H(V^n,X_1^n |U^n)+H(X_2^n|V^n,X_1^n,U^n) \ge n(R+I(X_1,X_2;Z|U)-R_2)+nR_2$ then
\begin{eqnarray}
\label{secrecy2}
nR_e &\ge& n(R+I(X_1,X_2;Z|U)-I(X_1,X_2;Z|U)-\epsilon_3)\notag\\
&=& n(R-\epsilon_3).
\end{eqnarray}
Therefore, perfect secrecy is obtained.

This completes the proof of Theorem 2.

\bibliographystyle{IEEEtran}
\bibliography{macsecrecy}

\begin{thebibliography}{10}
\providecommand{\url}[1]{#1}
\csname url@samestyle\endcsname
\providecommand{\newblock}{\relax}
\providecommand{\bibinfo}[2]{#2}
\providecommand{\BIBentrySTDinterwordspacing}{\spaceskip=0pt\relax}
\providecommand{\BIBentryALTinterwordstretchfactor}{4}
\providecommand{\BIBentryALTinterwordspacing}{\spaceskip=\fontdimen2\font plus
\BIBentryALTinterwordstretchfactor\fontdimen3\font minus
  \fontdimen4\font\relax}
\providecommand{\BIBforeignlanguage}[2]{{%
\expandafter\ifx\csname l@#1\endcsname\relax
\typeout{** WARNING: IEEEtran.bst: No hyphenation pattern has been}%
\typeout{** loaded for the language `#1'. Using the pattern for}%
\typeout{** the default language instead.}%
\else
\language=\csname l@#1\endcsname
\fi
#2}}
\providecommand{\BIBdecl}{\relax}
\BIBdecl

\bibitem{wyner}
A.~D. Wyner, ``The wiretap channel,'' \emph{Bell System Technical Journal},
  vol.~54, pp. 1355--1387, Oct. 1975.

\bibitem{leung}
S.~K. Leung-Yan-Cheong and M.~Hellman, ``The {G}aussian wire-tap channel,''
  \emph{IEEE Transactions on Information Theory}, vol.~24, no.~4, pp. 451--456,
  Jul. 1978.

\bibitem{csiszar}
I.~Csisz\'{a}r and J.~K{\"{o}}rner, ``Broadcast channels with confidential
  messages,'' \emph{IEEE Transactions on Information Theory}, vol.~24, no.~3,
  pp. 339--348, May. 1978.

\bibitem{liangshamai}
Y.~Liang, H.~V. Poor, and S.~{Shamai (Shitz)}, ``Secure communication over
  fading channels,'' \emph{IEEE Transactions on Information Theory}, vol.~54,
  no.~6, pp. 2470--2492, Jun. 2008.

\bibitem{khisti}
A.~Khisti and G.~W. Wornell, ``Secure transmission with multiple antennas {II}:
  The {MIMOME} wiretap channel,'' \emph{IEEE Transactions on Information
  Theory}, vol.~56, no.~11, pp. 5515--5532, Nov. 2010.

\bibitem{oggier}
F.~Oggier and B.~Hassibi, ``The secrecy capacity of the {MIMO} wiretap
  channel,'' \emph{IEEE Transactions on Information Theory}, vol.~57, no.~8,
  pp. 4961--4972, Aug. 2011.

\bibitem{Tie}
T.~Liu and S.~{Shamai (Shitz)}, ``A note on the secrecy capacity of the
  multi-antenna wiretap channel,'' \emph{IEEE Transactions on Information
  Theory}, vol.~55, no.~6, pp. 2547--2553, Jun. 2009.

\bibitem{tekin}
E.~Tekin and A.~Yener, ``The {G}aussian multiple access wire-tap channel,''
  \emph{IEEE Transactions on Information Theory}, vol.~54, no.~12, pp.
  5747--5755, Dec. 2008.

\bibitem{liangpoor}
Y.~Liang and H.~V. Poor, ``Multiple access channels with confidential
  messages,'' \emph{IEEE Transactions on Information Theory}, vol.~54, no.~3,
  pp. 976--1002, Mar. 2008.

\bibitem{tekin2}
E.~Tekin and A.~Yener, ``The general {G}aussian multiple access and two-way
  wire-tap channels: Achievable rates and cooperative jamming,'' \emph{IEEE
  Transactions on Information Theory}, vol.~54, no.~6, pp. 2735--2751, Jun.
  2008.

\bibitem{yuksel1}
M.~Yuksel and E.~Erkip, ``The relay channel with a wire-tapper,'' in \emph{41st
  Annual Conference on Information Sciences and Systems}, Baltimore, MD, USA,
  Mar. 2007, pp. 13--18.

\bibitem{yuksel2}
------, ``Secure communication with a relay helping the wire-tapper,'' in
  \emph{IEEE Information Theory Workshop}, Lake Tahoe, CA, USA, Sep. 2007, pp.
  595--600.

\bibitem{xianghe}
X.~He and A.~Yener, ``Cooperation with an untrusted relay: A secrecy
  perspective,'' \emph{IEEE Transactions on Information Theory}, vol.~56,
  no.~8, pp. 3801--3827, Aug. 2010.

\bibitem{lai}
L.~Lai and H.~E. Gamal, ``The relay eavesdropper channel: {C}ooperation for
  secrecy,'' \emph{IEEE Transactions on Information Theory}, vol.~54, no.~9,
  pp. 4005--4019, Sep. 2008.

\bibitem{vaneet}
V.~Aggarwal, L.~Sankar, A.~R. Calderbank, and H.~V. Poor, ``Secrecy capacity of
  a class of orthogonal relay eavesdropper channels,'' \emph{EURASIP Journal on
  Wireless Communications and Networking, Special Issue on Wireless Physical
  Layer Security}, 2009.

\bibitem{zohaibj}
Z.~H. Awan, A.~Zaidi, and L.~Vandendorpe, ``Secure communication over parallel
  relay channel,'' \emph{IEEE Transactions on Information Forensics and
  Security}, vol.~7, no.~2, pp. 359--371, Apr. 2012.

\bibitem{tang}
X.~Tang, R.~Liu, P.~Spasojevic, and H.~V. Poor, ``Interference assisted secret
  communication,'' \emph{IEEE Transactions on Information Theory}, vol.~57,
  no.~5, pp. 3153--3167, May. 2011.

\bibitem{onur}
O.~Koyluoglu and H.~El~Gamal, ``Cooperative encoding for secrecy in
  interference channels,'' \emph{IEEE Transactions on Information Theory},
  vol.~57, no.~9, pp. 5682 --5694, Sept. 2011.

\bibitem{ruoheng}
R.~Liu, I.~Maric, P.~Spasojevic, and R.~Yates, ``Discrete memoryless
  interference and broadcast channels with confidential messages: Secrecy rate
  regions,'' \emph{IEEE Transactions on Information Theory}, vol.~54, no.~6,
  pp. 2493 --2507, June 2008.

\bibitem{vilela}
J.~Vilela, M.~Bloch, J.~Barros, and S.~McLaughlin, ``Wireless secrecy regions
  with friendly jamming,'' \emph{IEEE Transactions on Information Forensics and
  Security}, vol.~6, no.~2, pp. 256 --266, june 2011.

\bibitem{barros}
J.~Barros and M.~R.~D. Rodrigues, ``Secrecy capacity of wireless channels,'' in
  \emph{IEEE International Symposium on Information Theory}, Seattle, USA, Jul.
  2006, pp. 356--360.

\bibitem{gopala}
P.~K. Gopala, L.~Lai, and H.~E. Gamal, ``On the secrecy capacity of fading
  channels,'' \emph{IEEE Transactions on Information Theory}, vol.~54, no.~12,
  pp. 4687--4698, Oct. 2008.

\bibitem{liangbook}
Y.~Liang, H.~V. Poor, and S.~{Shamai (Shitz)}, ``Information theoretic
  security,'' \emph{Foundations and Trends in Communications and Information
  Theory}, vol.~5, no. 4-5, pp. 355--580, 2009.

\bibitem{willems}
F.~M.~J. Willems, ``The discrete memoryless multiple access channel with
  partially cooperating encoders,'' \emph{IEEE Transactions on Information
  Theory}, vol.~29, no.~3, pp. 441--445, May. 1983.

\bibitem{wigger}
S.~I. Bross, A.~Lapidoth, and M.~Wigger, ``The {G}aussian {MAC} with
  conferencing encoders,'' in \emph{IEEE International Symposium on Information
  Theory}, Toronto, ON, Jul. 2008, pp. 2702--2706.

\bibitem{kim}
Y.-H. Kim, ``Coding techniques for primitive relay channels,'' in \emph{45th
  Annual Allerton Conference Communication, Control and Computing}, Monticello,
  IL, USA, Sept. 2007, pp. 129--135.

\bibitem{xainghe}
X.~He and A.~Yener, ``Providing secrecy with structured codes: Tools and
  applications to two-user gaussian channels,'' \emph{available online
  http://arxiv.org/abs/0907.5388}, 2009.

\bibitem{LYT08}
Z.~Li, R.~Yates, and W.~Trappe, ``Secrecy capacity of a class of one-sided
  interference channel,'' in \emph{IEEE Int. Symp. on Information Theory}, Jul.
  2008, pp. 379--383.

\bibitem{gallager}
R.~G. Gallager, \emph{Information theory and reliable communication}.\hskip 1em
  plus 0.5em minus 0.4em\relax New York:Wiley, 1968.

\bibitem{km}
A.~Khisti and G.~W. Wornell, ``Secure transmission with multiple antennas {I}:
  The {MISOME} wiretap channel,'' \emph{IEEE Transactions on Information
  Theory}, vol.~56, no.~7, pp. 3088--3104, Jul. 2010.

\bibitem{cover}
T.~M. Cover and J.~A. Thomas, \emph{Elements of Information Theory}.\hskip 1em
  plus 0.5em minus 0.4em\relax New York:Wiley, 1991.

\end{thebibliography}

\begin{IEEEbiography}[{\includegraphics[width=1.1in,height=1.3in]{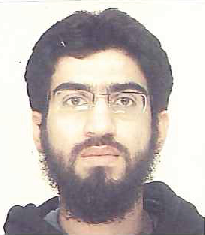}}]
{Zohaib Hassan Awan}  
 received the B.S. degree in Electronics Engineering from Ghulam Ishaq Khan Institute (GIKI), Topi, Pakistan in 2005 and the M.S. degree in Electrical Engineering with a majors in wireless systems from Royal Institute of Technology (KTH), Stockholm, Sweden in 2008. He received the Ph.D. degree in Electrical Engineering from Universit\'{e} catholique de Louvain (UCL), Belgium in 2013.

His current research interests include information-theoretic security, cooperative communications and communication theory.
\end{IEEEbiography}

\begin{IEEEbiography}[{\includegraphics[width=1.1in,height=1.3in]{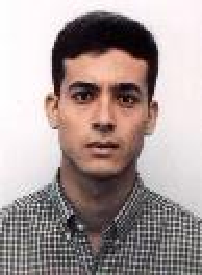}}]
{Abdellatif Zaidi}  
 received the B.S. degree in Electrical Engineering from \'{E}cole Nationale Sup\'{e}rieure de Techniques Avanc\'{e}s, ENSTA ParisTech, France in 2002 and the M. Sc. and Ph.D. degrees in Electrical Engineering from \'{E}cole Nationale Sup\'{e}rieure des T\'{e}l\'{e}communications, TELECOM ParisTech, Paris, France in 2002 and 2005, respectively.

From December 2002 to December 2005, he was with the Communications and Electronics Dept., TELECOM ParisTech, Paris, France and the Signals and Systems Lab., CNRS/Sup\'{e}lec, France pursuing his PhD degree. From May 2006 to September 2010, he was at \'{E}cole Polytechnique de Louvain, Universit\'{e} catholique de Louvain, Belgium, working as a research assistant. Dr. Zaidi was "Research Visitor" at the University of Notre Dame, Indiana, USA,
during fall 2007 and Spring 2008. He is now, an assistant professor at Universit\'e Paris-Est Marne-la-Vall\'ee, France.

His research interests cover a broad range of topics from signal processing for communication and multi-user information theory. Of particular interest are the problems of coding for side-informed channels, secure communication, coding and interference mitigation in multi-user channels, and relaying problems and cooperative communication with application to sensor networking and ad-hoc wireless networks.
\end{IEEEbiography}

\begin{IEEEbiography}[{\includegraphics[width=1.1in,height=1.3in]{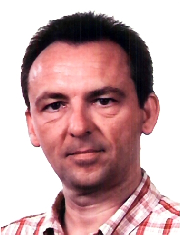}}]
{Luc Vandendorpe} (M'93-SM'99-F'06)
 was born in Mouscron, Belgium, in 1962. He received the Electrical Engineering degree (summa cum laude) and the Ph. D. degree from the Universit\'{e} catholique de Louvain (UCL) Louvain-la- Neuve, Belgium in 1985 and 1991 respectively. Since 1985, L. Vandendorpe is with the Communications and Remote Sensing Laboratory of UCL where he first worked in the field of bit rate reduction techniques for video coding. In 1992, he was a Visiting Scientist and Research Fellow at the Telecommunications and Traffic Control Systems Group of the Delft Technical University, Netherlands, where he worked on Spread Spectrum Techniques for Personal Communications Systems. From October 1992 to August 1997, L. Vandendorpe was Senior Research Associate of the Belgian NSF at UCL. Presently, he is Full Professor and Head of the Institute for Information and Communication Technologies, Electronics and Applied Mathematics of UCL.

His current interest is in digital communication systems and more precisely resource allocation for OFDM(A) based multicell systems, MIMO and distributed MIMO, sensor networks, turbo-based communications systems, physical layer security and UWB based positioning.

In 1990, he was co-recipient of the Biennal Alcatel-Bell Award from the Belgian NSF for a contribution in the field of image coding. In 2000 he was co-recipient (with J. Louveaux and F. Deryck) of the Biennal Siemens Award from the Belgian NSF for a contribution about filter bank based multicarrier transmission. In 2004 he was co-winner (with J. Czyz) of the Face Authentication Competition, FAC 2004. L. Vandendorpe is or has been TPC member for numerous IEEE conferences (VTC Fall, Globecom Communications Theory Symposium, SPAWC, ICC) and for the Turbo Symposium. He was co-technical chair (with P. Duhamel) for IEEE ICASSP 2006.

He was an editor of the IEEE Trans. on Communications for Synchronization and Equalization between 2000 and 2002, associate editor of the IEEE Trans. on Wireless Communications between 2003 and 2005, and associate editor of the IEEE Trans. on Signal Processing between 2004 and 2006. He was chair of the IEEE Benelux joint chapter on Communications and Vehicular Technology between 1999 and 2003. He was an elected member of the Signal Processing for Communications committee between 2000 and 2005, and between 2009 and 2011, and an elected member of the Sensor Array and Multichannel Signal Processing committee of the Signal Processing Society between 2006 and 2008. Currently, he is the Editor in Chief for the Eurasip Journal on Wireless Communications and Networking. L. Vandendorpe is a Fellow of the IEEE.
\end{IEEEbiography}

\end{document}